# Chaotic Scattering Theory, Thermodynamic Formalism, and Transport Coefficients

P. Gaspard

*Faculté des Sciences and Centre for Nonlinear Phenomena and Complex Systems, Université Libré de Bruxelles, Campus Plaine, Code Postal 231, B-1050, Brussels, Belgium*

J.R. Dorfman

(April 26, 1995)

The foundations of the chaotic scattering theory for transport and reaction-rate coefficients for classical many-body systems are considered here in some detail. The thermodynamic formalism of Sinai, Bowen, and Ruelle is employed to obtain an expression for the escape rate for a phase space trajectory of a system to leave a finite open region of phase space for the first time. This expression relates the escape rate to the difference between the sum of the positive Lyapunov exponents and the Kolmogorov-Sinai entropy for the fractal set of phase space trajectories which are trapped forever in the open region. This relation is well known for systems of a few degrees of freedom, and is here extended to systems with many degrees of freedom. The formalism is applied to smooth hyperbolic systems, to cellular-automata lattice gases, and to hard-sphere systems. In the latter case, the geometric constructions of Sinai *et al.* for billiard systems are used to describe the relevant chaotic scattering phenomena. Some applications of this formalism to nonhyperbolic systems are also discussed.

## I. INTRODUCTION

In a previous paper (I) [1], we extended the chaotic scattering, or escape-rate method of Gaspard and Nicolis [2] for the coefficient of diffusion of the moving particle in Lorentz gas systems so as to apply to a wider class of transport coefficients for a simple, classical fluid and to chemical reaction rates. The line of the argument in I was as follows: (1) One can associate to every transport process in a fluid a microscopic, dynamical quantity called a Helfand moment [3]. (2) For large enough systems and for long enough times, the mean square fluctuations of the Helfand moments about their initial values grow linearly with the time $t$, in the case that normal hydrodynamic processes take place in the fluid. The coefficient of the linear term in $t$ for each Helfand moment is, apart from numerical factors, the relevant transport coefficient. (3) This "diffusion" of the Helfand moment can be regarded as the result of chaotic scattering processes in an appropriate phase space. The dynamics of the Helfand moments become more diffusion-like as the system size increases and as the time becomes longer, due to the occurrence of more and more individual scattering events taking place as the time gets longer. (4) One can characterize this diffusion-like process in phase space in terms of an escape rate of trajectories from regions where the Helfand moment lies within some prescribed bounds into regions where the value lies outside these bounds. This is the phase space analog of characterizing Brownian motion by the rate at which Brownian particles pass into an absorbing boundary. (5) One then relates this escape rate for a Helfand moment to a transport coefficient on one hand, and on the other hand, to the sum of the positive Lyapunov exponents and to the Kolmogorov-Sinai (KS) entropy that characterizes the set of trajectories in phase space where the Helfand moment lies forever within prescribed bounds. These trajectories form an unstable fractal set in phase space, a fractal repeller, denoted by $\mathcal{R}_\chi^{(\alpha)}$, where $\alpha$ denotes the transport coefficient associated with a time dependent Helfand moment $G_t^{(\alpha)}$, which on the fractal repeller remains within the bounds

$$-\chi/2 \leq G_t^{(\alpha)} \leq +\chi/2 \qquad (1)$$

where $\chi$ is real and positive.

The main result of I is that

$$\alpha = \lim_{\chi \to \infty} \left(\frac{\chi}{\pi}\right)^2 \lim_{V \to \infty} \Big[\sum_{\lambda_i > 0} \lambda_i(\mathcal{R}_\chi^{(\alpha)}) - h_{\text{KS}}(\mathcal{R}_\chi^{(\alpha)})\Big] \qquad (2)$$

where $\sum_{\lambda_i>0} \lambda_i(\mathcal{R}_\chi^{(\alpha)})$ is the sum over all positive Lyapunov exponents for the trajectories on the fractal repeller, and $h_{\text{KS}}(\mathcal{R}_\chi^{(\alpha)})$ is the Kolmogorov-Sinai entropy for trajectories on the repeller. Here the limit $V \to \infty$ denotes the thermodynamic limit, taken before the limit $\chi \to \infty$.

The purpose of this paper is to discuss the chaotic scattering theory which leads to the relation between the escape rate $\gamma_\chi^{(\alpha)}$ of the Helfand moment from the region (1), and the dynamical quantities $\lambda_i(\mathcal{R}_\chi^{(\alpha)})$ and $h_{\mathrm{KS}}(\mathcal{R}_\chi^{(\alpha)})$, that is

$$\gamma_\chi^{(\alpha)} = \sum_{\lambda_i > 0} \lambda_i(\mathcal{R}_\chi^{(\alpha)}) - h_{\mathrm{KS}}(\mathcal{R}_\chi^{(\alpha)}). \tag{3}$$

This result is already well known, due to the work of Kantz and Grassberger [4], Eckmann and Ruelle [5], Bohr and Rand [6], Tel and co-workers [7], Grebogi, Ott, and Yorke [8], and Kadanoff and Tang [9], for systems with a few degrees of freedom. Our goal here is to extend these previous discussions to systems with many degrees of freedom, and to provide the necessary foundations for the application of this method to a number of systems of physical interest. These systems include smooth, hyperbolic as well as some nonhyperbolic systems, cellular-automata lattice gases, and systems of hard-sphere particles. In the latter case a more delicate analysis of phase space trajectories than that needed for smooth systems must be carried out due to the discontinuous nature of the hard-sphere potential.

Our work will, in the main, draw upon two mathematical developments, both stimulated by problems of interest to statistical mechanics. The analysis of the connection between chaotic scattering and escape rates presented here, like that of some other workers [6,7], is based upon the thermodynamic formalism for describing dynamical systems [10–13]. In this analysis, the escape-rate formula, Eq. (3), is a consequence of the properties of the Ruelle, or topological, pressure that arises naturally in the thermodynamic formalism. Our principal contribution here is to present the method in a way that it applies to systems with many degrees of freedom, to emphasize some of the mathematical literature that has a direct bearing on problems of physical interest, and to show how the method can be applied to some systems of current physical interest, including some nonhyperbolic systems.

The other mathematical development which we will use is the analysis of the dynamical properties of hard-sphere systems based on a "ray optics" description given by Sinai, Bunimovich, Chernov and others [14,15]. This description involves a careful study of the differential geometry of the phase space trajectories for hard-sphere systems and leads to a continued fraction expansion of the so-called second fundamental operator. This result is of deep fundamental interest for describing the ergodic properties of hard-sphere systems. The second fundamental operator can be used to compute quantities such as Lyapunov exponents and Kolmogorov-Sinai entropies for such systems. An important application of these and related methods to the triangular Lorentz gas has recently been given by Gaspard and Baras [16,17]. Furthermore, van Beijeren and Dorfman [18] have shown that there is a clear connection between the differential geometry analysis of Sinai *et al.* and the kinetic theory of gases. This leads us to believe that it should be possible to reformulate kinetic theory in such a way that its reliance on the chaotic behavior of gases is clearly well founded upon a mathematical analysis of the underlying dynamics rather than assumed, as is usually done. This task awaits further study and will not be considered here.

The plan of this paper is as follows. In section II, we present the thermodynamic formalism for smooth hyperbolic systems, derive the escape-rate formula from the properties of the Ruelle pressure, and extend the analysis to nonhyperbolic systems. In section III, we apply the geometric methods of Sinai *et al.* to describe the dynamical properties of hard-sphere systems in terms of the second fundamental form. In section IV, we show that the ideas of the thermodynamic formalism can easily be applied to cellular-automata lattice gases, which are of interest as simple models that exhibit hydrodynamic phenomena typical of real fluids. This topic is explored in considerable detail in related work by Ernst, Dorfman, Nix, and Jacobs [19]. In section V we present a brief discussion of related methods for describing transport phenomena in terms of the underlying dynamical properties of the system, and we conclude in section VI with a number of remarks outlining open problems and directions for further work.

## II. LARGE-DEVIATION OR THERMODYNAMIC FORMALISM FOR DYNAMIC INSTABILITIES

In this section we apply the large-deviation, or thermodynamic, formalism of Bowen, Ruelle, and Sinai [10–13] to study the statistical properties of the phase space trajectories of systems with smooth potential energies and which obey Hamiltonian mechanics. The purpose of this formalism is to use statistical properties of the trajectories in order to construct invariant probability measures. These measures in turn can be used to compute phase space averages which are needed for computing effects of the chaotic scattering phenomena of physical interest here. The large-deviation formalism which we use goes beyond a linear or quadratic deviation of some quantity from its reference value. For example, the chaotic behavior of a system depends upon *exponential* separation of trajectories in phase space. The large-deviation formalism is designed to treat such circumstances.

## A. Linear Stability and Lyapunov Exponents

### 1. The phase space and the tangent space

Consider a mechanical system of $N$ particles with Hamiltonian $H(\mathbf{q}, \mathbf{p})$ where $(\mathbf{q}, \mathbf{p})$ is a $2Nf$ dimensional vector space with $\mathbf{q} = (\mathbf{q}_1, \mathbf{q}_2, \cdots, \mathbf{q}_N)$ where $\mathbf{q}_i$ is the ($f$-dimensional) coordinate of particle $i$, $\mathbf{p} = (\mathbf{p}_1, \mathbf{p}_2, \cdots, \mathbf{p}_N)$ and $\mathbf{p}_i$ is the ($f$-dimensional) momentum of particle $i$. The Hamiltonian function has the form

$$H(\mathbf{q}, \mathbf{p}) = \sum_{i=1}^{N} \frac{\mathbf{p}_i^2}{2m} + V(\mathbf{q}) \tag{4}$$

where $V(\mathbf{q})$ is the potential energy of interaction between the particles. To avoid bound states, and orbiting collisions, we assume that the potential energy is a sum of central, short range, repulsive pair potentials. Hamilton's equations of motion are

$$\dot{\mathbf{q}} = \frac{\partial H(\mathbf{q}, \mathbf{p})}{\partial \mathbf{p}}, \quad \dot{\mathbf{p}} = -\frac{\partial H(\mathbf{q}, \mathbf{p})}{\partial \mathbf{q}} \tag{5}$$

We will suppose either that the boundaries of the system are hard walls with infinite mass whose shape will be discussed later, or that periodic boundary conditions are applied. In both cases the total energy is conserved, and in the case of periodic boundary conditions, the total momentum is also conserved. We will always consider a system at one fixed energy $E$, and therefore the $2Nf$ dimensional phase space reduces to the $2Nf - 1$ dimensional constant energy surface $\mathcal{M}$ defined by the condition that $H = E$. A point on the surface will be denoted by $\mathbf{X}$. The trajectory of a phase-space point, initially at $\mathbf{X}_0$, will be indicated by a flow $\Phi^t$ such that the phase-space point at a time $t$ later is given as $\mathbf{X}_t = \Phi^t \mathbf{X}_0$. Finally we denote by $\mathcal{TM}(\mathbf{X})$ a linear vector space which is tangent to $\mathcal{M}$ at the phase-space point $\mathbf{X}$ on the constant energy surface.

### 2. The fundamental matrix and its decomposition

An important characterization of the trajectories is given by their stability and, in particular, by their linear stability which controls the way infinitesimal perturbations evolve with time in the tangent space. These perturbations can be calculated by integration of a coupled set of equations, one for the trajectory that passes through the point $\mathbf{X}$,

$$\dot{\mathbf{X}} = \mathbf{F}(\mathbf{X}) \tag{6}$$

which is a simple rewriting of Hamilton's equation, Eq. (5), and one for the a trajectory that deviates by an infinitesimal amount from the reference trajectory, $\mathbf{X}_t = \Phi^t \mathbf{X}_0$,

$$\delta \dot{\mathbf{X}} = \mathbf{F}(\mathbf{X} + \delta \mathbf{X}) - \mathbf{F}(\mathbf{X}) = \frac{\partial \mathbf{F}(\mathbf{X})}{\partial \mathbf{X}} \cdot \delta \mathbf{X} \tag{7}$$

to linear order in $\delta \mathbf{X}$. Since Eq. (7) is linear, all of its solutions can be expressed as

$$\delta \mathbf{X} = \Phi^t(\mathbf{X}_0 + \delta \mathbf{X}_0) - \Phi^t(\mathbf{X}_0) = \frac{\partial \Phi^t(\mathbf{X}_0)}{\partial \mathbf{X}_0} \cdot \delta \mathbf{X}_0 = \mathsf{M}(t, \mathbf{X}_0) \cdot \delta \mathbf{X}_0 \tag{8}$$

with

$$\mathsf{M}(t, \mathbf{X}_0) = \mathbb{T} \, \exp \int_0^t \frac{\partial \mathbf{F}}{\partial \mathbf{X}}(\Phi^\tau \mathbf{X}_0) \, d\tau \tag{9}$$

where a time-ordered exponential is indicated. The fundamental matrix thus obeys the evolution equation

$$\dot{\mathsf{M}}(t, \mathbf{X}_0) = \frac{\partial \mathbf{F}}{\partial \mathbf{X}}(\Phi^t \mathbf{X}_0) \cdot \mathsf{M}(t, \mathbf{X}_0) \tag{10}$$

In Eq. (8) we have set $\mathbf{X}_0$ and $\delta \mathbf{X}_0$ to be the value of $\mathbf{X}$ and $\delta \mathbf{X}$ at $t = 0$, and $\mathsf{M}(t, \mathbf{X}_0)$ is the fundamental matrix solution of Eq. (7). We mention for use in subsection II.D that the construction of the fundamental matrix requires us to follow two nearby trajectories through their full evolution over a time interval of duration $t$. Our primary interest resides in characterizing the rate of separation or of approach of two nearby trajectories in terms of characteristic exponents, called Lyapunov exponents. In a stability analysis, stable and unstable directions are determined by the signs of the Lyapunov exponents in an obvious way.

The Multiplicative Ergodic Theorem of Oseledets [20–22] allows us to clearly identify the locally stable and unstable directions. This is accomplished through the use of the property of the matrix $\mathsf{M}(t, \mathbf{X}_0)$ as a multiplicative cocycle. That is, $\mathsf{M}(t, \mathbf{X}_0)$ satisfies the important group relation

$$\mathsf{M}(t+s, \mathbf{X}) = \mathsf{M}(t, \Phi^s \mathbf{X}) \cdot \mathsf{M}(s, \mathbf{X}) \ , \tag{11}$$

for any positive time $s$. This is just a restatement of the time evolution property of the solution of Eq. (7). Next we define a Lyapunov homology as a local, linear transformation between the cocycle $\mathsf{M}$ and another cocycle $\mathsf{m}$ of the form

$$\mathsf{M}(t, \mathbf{X}) = \mathsf{C}(\Phi^t \mathbf{X}) \cdot \mathsf{m}(t, \mathbf{X}) \cdot \mathsf{C}^{-1}(\mathbf{X}) \tag{12}$$

such that the transformation matrices $\mathsf{C}$ have no exponential time dependence, that is, such that

$$\lim_{t \to \infty} \frac{1}{t} \ln \left\| \mathsf{C}(\Phi^t \mathbf{X}) \right\| = 0. \tag{13}$$

One can easily check that the cocycle $\mathsf{M}$ satisfies Eq. (11) provided that the cocycle $\mathsf{m}$ does also. The purpose of the Lyapunov homology is to find conditions under which the cocycle $\mathsf{M}$ might be reduced to a diagonal cocycle $\mathsf{m}$ such that

$$\mathsf{M}(t, \mathbf{X}) = \sum_k \mathbf{e}_k(\Phi^t \mathbf{X}) \Lambda_k(t, \mathbf{X}) \mathbf{f}_k^{\mathrm{T}}(\mathbf{X}) \tag{14}$$

where $\mathbf{e}_k(\mathbf{X})$ are vectors with components $e_{k,i}(\mathbf{X}) = C_{ik}(\mathbf{X})$ and $\mathbf{f}_k$ are vectors with components $f_{k,j}(\mathbf{X}) = [\mathsf{C}^{-1}(\mathbf{X})]_{jk}$. Here the superscript T denotes a transpose. These vectors form a set of bi-orthogonal pairs satisfying

$$\sum_k \mathbf{e}_k \, \mathbf{f}_k^{\mathrm{T}} = \mathbf{1} \ , \qquad \text{and} \qquad \mathbf{f}_k^{\mathrm{T}} \cdot \mathbf{e}_l = \delta_{kl} \tag{15}$$

The functions $\Lambda_k(t, \mathbf{X})$ are called stretching factors in the directions $\{\mathbf{e}_k\}$. They must also satisfy a multiplicative cocycle relation

$$\Lambda_k (t+s, \mathbf{X}) = \Lambda_k (t, \Phi^s \mathbf{X}) \, \Lambda_k (s, \mathbf{X}). \tag{16}$$

We emphasize that this construction is not a diagonalization of the matrix $\mathsf{M}$ in the usual sense because the vector $\mathbf{f}_k$ is evaluated at the initial point $\mathbf{X}$ while the vector $\mathbf{e}_k$ is evaluated at the final point $\Phi^t \mathbf{X}$. It is also important to note that $\mathbf{e}_k(\Phi^t \mathbf{X})$ and $\mathbf{f}_k(\mathbf{X})$ are not, in general, mutually orthogonal. The decomposition (14) of the fundamental matrix is also referred to as the Mather spectrum [23].

When the matrix $\mathsf{M}$ is applied to one of the direction vectors $\mathbf{e}_k$ one obtains

$$\mathsf{M}(t, \mathbf{X}) \cdot \mathbf{e}_k (\mathbf{X}) = \Lambda_k (t, \mathbf{X}) \, \mathbf{e}_k (\Phi^t \mathbf{X}) \tag{17}$$

According to the Lyapunov condition Eq. (13), the vector $\mathbf{e}_k(\Phi^t \mathbf{X})$ has no exponential time dependence, so that the entire exponential behavior, if any, on the right hand side of Eq. (17) is contained in the function $\Lambda_k (t, \mathbf{X})$. Since the Lyapunov exponents measure the rate of exponential separation or of approach of two nearby trajectories in different directions in phase space, we see that the Lyapunov exponent associated with the direction $\mathbf{e}_k (\mathbf{X})$ is

$$\lambda_k (\mathbf{X}) = \lim_{t \to \infty} \frac{1}{t} \ln |\Lambda_k(t, \mathbf{X})| \tag{18}$$

If the initial perturbation, $\delta \mathbf{X}_0 = \bar{\mathbf{e}}$, points toward an arbitrary direction in the tangent space, its time evolution will be determined by the largest among the stretching factors for which the scalar products $\mathbf{f}_k^{\mathrm{T}} \cdot \bar{\mathbf{e}}$ do not nonvanish as we can conclude from the relation (8) defining the fundamental matrix and its decomposition (14).

In this way Oseledets' theorem shows that the Lyapunov exponent associated with an arbitrary tangent vector $\bar{\mathbf{e}}$,

$$\lambda (\mathbf{X}, \bar{\mathbf{e}}) = \lim_{t \to \infty} \frac{1}{t} \ln \| \mathsf{M}(t, \mathbf{X}) \cdot \bar{\mathbf{e}} \| \tag{19}$$

takes its value from a discrete set called the spectrum of Lyapunov exponents which satisfy

$$\lambda^{(1)}(\mathbf{X}) > \lambda^{(2)}(\mathbf{X}) > \cdots > \lambda^{(r)}(\mathbf{X}) \tag{20}$$

with multiplicities

$$m^{(1)}(\mathbf{X}), m^{(2)}(\mathbf{X}), ..., m^{(r)}(\mathbf{X}) \tag{21}$$

which sum up to the dimension of the tangent space $M = \dim \mathcal{TM} = \dim \mathcal{M}$

$$M = \sum_{n=1}^{r} m^{(n)}(\mathbf{X}) = 2Nf - 1 \tag{22}$$

The value $\lambda^{(n)}(\mathbf{X})$ is obtained from Eq. (19) when the tangent vector $\bar{\mathbf{e}}$ belongs to the subspace $\mathcal{V}^{(n)}(\mathbf{X}) \setminus \mathcal{V}^{(n+1)}(\mathbf{X})$ (that is, the set of points in $\mathcal{V}^{(n)}$ but not in $\mathcal{V}^{(n+1)}$), where $\{\mathcal{V}^{(n)}(\mathbf{X})\}_{n=1}^{r}$ are nested linear subspaces of the tangent space

$$\mathcal{TM}(\mathbf{X}) = \mathcal{V}^{(1)}(\mathbf{X}) \supset \mathcal{V}^{(2)}(\mathbf{X}) \supset \cdots \supset \mathcal{V}^{(r)}(\mathbf{X}),$$

such that $m^{(n)}(\mathbf{X}) = \dim[\mathcal{V}^{(n)}(\mathbf{X}) \setminus \mathcal{V}^{(n+1)}(\mathbf{X})]$. The linear subspace $\mathcal{V}^{(n)}(\mathbf{X})$ is spanned by the set of unit vectors $\{\mathbf{e}_i(\mathbf{X})\}_{i \in I^{(n)}}$ such that the corresponding Lyapunov exponents in Eq. (20) satisfy $\lambda_i(\mathbf{X}) \leq \lambda^{(n)}(\mathbf{X})$ for $i \in I^{(n)}$.

## 3. Stable and unstable manifolds

The stable and unstable manifolds $W_s(\mathbf{X}_0)$ and $W_u(\mathbf{X}_0)$, respectively, associated with the trajectory at $\mathbf{X}_0$ play a central role in dynamical systems theory [5,23–25]. These manifolds are defined by

$$W_s(\mathbf{X}_0) = \{\mathbf{X} \in \mathcal{M} : \|\Phi^t \mathbf{X} - \Phi^t \mathbf{X}_0\| \to 0 \quad \text{for} \quad t \to +\infty\} \tag{23}$$

and

$$W_u(\mathbf{X}_0) = \{\mathbf{X} \in \mathcal{M} : \|\Phi^t \mathbf{X} - \Phi^t \mathbf{X}_0\| \to 0 \quad \text{for} \quad t \to -\infty\} \tag{24}$$

where $\|\cdot\|$ denotes a distance which we may take to be a Riemannian metric distance on the constant energy surface. The stable and unstable manifolds are global objects extending in phase space, as illustrated in Fig. 1. When restricted to the vicinity of the trajectory, these manifolds are referred to as the local stable and local unstable manifolds. The local stable and unstable manifolds are tangent respectively to the stable and unstable directions given by the corresponding vector fields $\{\mathbf{e}_k(\mathbf{X})\}$ of the tangent space. The union of all the stable or unstable manifolds of all the points of a trajectory $\cup_{-\infty < t < +\infty} W_{s,u}(\Phi^t \mathbf{X}_0)$ are invariant under the time evolution.

The concept of stable and unstable manifolds allow us to obtain an alternative method of calculation of the Lyapunov exponents. If, in the definition of the global invariant manifold, Eqs. (23) and (24) we replace $\mathbf{X}$ by $\mathbf{X}_0 + \delta \mathbf{X}_0$ we obtain the definitions of the local stable and unstable manifolds in terms of the distance

$$\|\delta \mathbf{X}_t\|^2 = \delta \mathbf{X}_0^{\mathrm{T}} \cdot \mathsf{M}^{\mathrm{T}}(t, \mathbf{X}_0) \cdot \mathsf{M}(t, \mathbf{X}_0) \cdot \delta \mathbf{X}_0, \tag{25}$$

which defines a positive definite quadratic form. We remark that this quadratic form is strictly internal to the tangent space at the only initial point $\mathbf{X}_0$, so that there is no explicit reference to the infinitesimal vector $\delta \mathbf{X}_t$ at another point on the trajectory. As a consequence, the structure of the stable and unstable manifolds, as well as the problem of linear stability, can be solved locally at each point $\mathbf{X}_0$ of the flow. Of course, the stability problem requires an integration of the linearized trajectory in the tangent plane to the constant energy surface at $\mathbf{X}_0$ and this information is contained in $\mathsf{M}^{\mathrm{T}}(t, \mathbf{X}_0)$ and $\mathsf{M}(t, \mathbf{X}_0)$ in Eq. (25).

The quadratic form in Eq. (25), $\mathsf{M}^{\mathrm{T}} \cdot \mathsf{M}$, can be diagonalized in terms of eigenvalues $\{\sigma_i(t, \mathbf{X}_0)\}$, and orthonormal eigenvectors $\{\mathbf{u}_i(t, \mathbf{X}_0)\}$ as

$$\mathsf{M}^{\mathrm{T}}(t, \mathbf{X}_0) \cdot \mathsf{M}(t, \mathbf{X}_0) = \sum_{i=1}^{2Nf-1} \mathbf{u}_i^{\mathrm{T}}(t, \mathbf{X}_0) \sigma_i(t, \mathbf{X}_0) \mathbf{u}_i(t, \mathbf{X}_0) \tag{26}$$

where we have used the fact that the eigenvectors must span the $2Nf-1$ dimensional space $\mathcal{TM}(\mathbf{X}_0)$ tangent to the constant energy surface at $\mathbf{X}_0$. The local Lyapunov exponents of the trajectory at initial point $\mathbf{X}_0$ may now be defined as

$$\lambda_i(\mathbf{X}_0) = \lim_{t\to\infty} \frac{1}{2t} \ln \sigma_i(t, \mathbf{X}_0). \tag{27}$$

which gives results which are equivalent to Eq. (18). Depending on the sign of the Lyapunov exponent $\lambda_i(\mathbf{X}_0)$, the corresponding directions are stable ($\lambda_i < 0$), or unstable ($\lambda_i > 0$), or neutral ($\lambda_i = 0$). The neutral directions include the direction of the flow (and, if we wished to include it, the direction perpendicular to the energy surface). We note that the orthonormal eigenvectors $\mathbf{u}_i(t, \mathbf{X}_0)$ are not directly related to the directions $\{\mathbf{e}_k(\mathbf{X}_0)\}$ which are not orthogonal. As a result of these considerations, we are able to express the tangent space at $\mathbf{X}_0$ as a direct sum of three linear subspaces

$$\mathcal{TM}(\mathbf{X}_0) = \mathcal{E}_u(\mathbf{X}_0) \oplus \mathcal{E}_0(\mathbf{X}_0) \oplus \mathcal{E}_s(\mathbf{X}_0), \tag{28}$$

spanned by the unstable, the neutral, and the stable directions, respectively, by merging together the linear subspaces $\mathcal{V}^{(n)}(\mathbf{X}_0) \setminus \mathcal{V}^{(n+1)}(\mathbf{X}_0)$ with $\lambda^{(n)} > 0$, $\lambda^{(n)} = 0$, and $\lambda^{(n)} < 0$, respectively.

### 4. Local stretching rates

We notice that the local positive Lyapunov exponents are quantities which are constant on half-trajectories $\Phi^t \mathbf{X}$ with $t \in [0, +\infty)$ or $(-\infty, 0]$ in the sense that $\lambda_k(\mathbf{X}) = \lambda_k(\Phi^s \mathbf{X})$ for $s \in \mathbb{R}^{\pm}$, which follows from the cocycle property (16) and the definition (18).

Our purpose is here to introduce the local stretching rates which are underlying the local Lyapunov exponents and which could be varying functions along each trajectory. With this aim, we differentiate Eq. (17) with respect to time to get an equation of evolution for the stretching factors $\Lambda_k(t, \mathbf{X})$. Using the evolution equation (10) of the fundamental matrix and the bi-orthogonality (15), we obtain

$$\dot{\Lambda}_k(t, \mathbf{X}) = \chi_k(\Phi^t \mathbf{X}) \Lambda_k(t, \mathbf{X}), \qquad \text{or} \qquad \Lambda_k(t, \mathbf{X}) = \exp \int_0^t \chi_k(\Phi^\tau \mathbf{X}) d\tau \tag{29}$$

where we introduced the *local stretching rate* in the direction $\mathbf{e}_k(\mathbf{X})$ by[1]

$$\chi_k(\mathbf{X}) = \mathbf{f}_k^{\mathrm{T}}(\mathbf{X}) \cdot \frac{\partial \mathbf{F}}{\partial \mathbf{X}}(\mathbf{X}) \cdot \mathbf{e}_k(\mathbf{X}) - \mathbf{f}_k^{\mathrm{T}}(\mathbf{X}) \cdot \frac{\partial \mathbf{e}_k}{\partial \mathbf{X}}(\mathbf{X}) \cdot \mathbf{F}(\mathbf{X}) \tag{30}$$

These quantities are defined locally at each phase-space point. Accordingly, the local Lyapunov exponents are given by

$$\lambda_k(\mathbf{X}) = \lim_{T\to\infty} \frac{1}{T} \int_0^T \chi_k(\Phi^t \mathbf{X}) dt = \lim_{T\to\infty} \frac{1}{2T} \int_{-T}^{+T} \chi_k(\Phi^t \mathbf{X}) dt \tag{31}$$

where we used the time reversibility. We remark that, for linear vector fields, $\mathbf{F}(\mathbf{X})$, the first term of Eq. (30) directly provides the local Lyapunov exponent $\lambda_k$. The linear stability analysis shows therefore that all of the Lyapunov exponents are defined in terms of the local stretching rates $\chi_k(\mathbf{X})$. The sum of the local stretching rates

$$u(\mathbf{X}) = \sum_{\lambda_k > 0} \chi_k(\mathbf{X}) \tag{32}$$

defines at each phase-space points $\mathbf{X}$ the quantity we call the *local dispersion rate*. This quantity plays a central role in the following considerations. Let us note that some of the local stretching factors may be locally negative although the corresponding local Lyapunov exponent is positive so that the sum should extend over all the positive local Lyapunov exponents.

Furthermore, we have the approximate relation

$$\prod_{\sigma_i > 1} \sqrt{\sigma_i(T, \mathbf{X})} \simeq \prod_{\Lambda_k > 1} \Lambda_k(T, \mathbf{X}) = \exp\left[ \int_0^T \sum_{\lambda_k > 0} \chi_k(\Phi^t \mathbf{X}) \, dt \right] \quad \text{as } T \to \infty. \tag{33}$$

which will be used below.

---

[1] These local stretching rates do not seem to have already been introduced in the literature.

## B. Symplectic dynamics, the pairing rule of the Lyapunov exponents, and hyperbolicity

Since Hamiltonian systems are symplectic, the fundamental matrix $\mathbf{M}(t, \mathbf{X}_0)$ obeys the relation

$$\mathbf{M}^T \cdot \mathbf{\Sigma} \cdot \mathbf{M} = \mathbf{\Sigma}$$

with the symplectic form $\mathbf{\Sigma}$ given by

$$\mathbf{\Sigma} = \begin{pmatrix} \mathbf{0} & \mathbf{1} \\ -\mathbf{1} & \mathbf{0} \end{pmatrix}. \tag{34}$$

As a result of this relation, if $\sigma$ is an eigenvalue of $\mathbf{M}^T \cdot \mathbf{M}$ then so is $\sigma^{-1}$. Accordingly, to each stable direction, there corresponds an unstable direction, and *vice versa*. Their Lyapunov exponents are respectively $-\lambda_i$ and $+\lambda_i$. This pairing rule also implies that the sum of all of the Lyapunov exponents must be zero for conservative Hamiltonian systems and that phase-space volumes are preserved by the dynamics.

In the case that the Hamiltonian system has periodic orbits, the linear stability of these orbits can be characterized by the eigenvalues $\Lambda_i$ of the fundamental matrix $\mathbf{M}(T)$ where $T$ is the period of the orbit. Indeed, the $\Lambda_i$ are related to the preceding eigenvalues by $\sigma_i = |\Lambda_i|^2$, so that

$$\lambda_i = \frac{1}{T} \ln |\Lambda_i|. \tag{35}$$

It is worth noting that to determine the Lyapunov exponents for non-periodic points $\mathbf{X}_0$, one should, in general determine the Lyapunov exponents from the eigenvalues of the symmetric, positive definite form $\mathbf{M}^T(t, \mathbf{X}_0) \cdot \mathbf{M}(t, \mathbf{X}_0)$ but for a point $\mathbf{X}$ on a periodic orbit, one may determine the Lyapunov exponents by computing the eigenvalues of $\mathbf{M}(T, \mathbf{X})$ directly.

A dynamical system is said to be *hyperbolic* if all the periodic orbits are unstable of saddle type with non-vanishing Lyapunov exponents (except for the zero exponents associated with the direction of flow as well as with any known, globally conserved quantities).[2] We shall say that the system is *nonhyperbolic* if there is a set of periodic orbits for which all of the Lyapunov exponents are zero. We will consider systems of both hyperbolic and nonhyperbolic types, although there are only a few general statements that can be made about nonhyperbolic systems.

## C. The Large-Deviation Formalism

The central quantity of interest in the large-deviation, or thermodynamic, formalism is an invariant probability measure on the set of phase-space trajectories. Here we outline the construction of the probability measure and define a fundamental quantity, the topological pressure. It will be important to keep in mind both the methods and results of equilibrium statistical mechanics of classical systems, since this subject motivates a large part of the discussion to follow, as was first demonstrated by Sinai, Ruelle, and Bowen [10–13]. In this context, the thermodynamic formalism is considered in time instead of space in contrast with equilibrium statistical mechanics. The topological pressure plays a role in dynamical systems theory very similar to that of the free energy for statistical mechanical systems. In addition it also is essential for establishing an important connection between the invariant probability measure on trajectories and the microcanonical measure on the constant energy surface. This connection lies at the heart of the large-deviation formalism, which, following the methods and ideas of Sinai, Ruelle, and Bowen, we hope to make clear below.

### 1. Separated subsets and topological entropy

We begin by considering the neighborhood of a point $\mathbf{X}_0$, and suppose that some of the Lyapunov exponents, $\lambda_i(\mathbf{X}_0)$ are positive. Then a typical neighborhood of this point of radius $\varepsilon$ say, will in the course of time be exponentially stretched in some directions and exponentially squeezed in others, while the phase-space measure of this small region remains constant in time. We will want to examine a very small subregion of this neighborhood such that *all* trajectories starting from points in the small subregion will remain within a distance $\varepsilon$ of the trajectory from $\mathbf{X}_0$ at any time over a time interval $(-T, +T)$. Since trajectories separate exponentially rapidly from each other, the size of this subregion must be exponentially small, with dimension of order

$$\varepsilon \exp\left[-\int_{-T}^{+T} \chi_1(\Phi^t \mathbf{X}_0) dt\right] \tag{36}$$

where $\chi_1$ is the maximum of the local stretching rates (30).

---

[2] We do not include in the definition of hyperbolicity the condition that the stable and unstable directions vary continuously with the phase-space point. This condition does not hold for dispersing billiards.

We make this argument more precise and generalize it somewhat as follows. We use the notion of $(\varepsilon, T)$-separated subsets of Bowen and Walters [11,26]. Consider a time $T$ and *define* a new distance between two points $\mathbf{X}_1$ and $\mathbf{X}_2$, $\rho_T(\mathbf{X}_1, \mathbf{X}_2)$, on the constant energy surface by

$$\rho_T(\mathbf{X}_1, \mathbf{X}_2) = \max_{-T \leq t \leq +T} \left\| \Phi^t \mathbf{X}_1 - \Phi^t \mathbf{X}_2 \right\|. \tag{37}$$

If it were to happen that $\rho_T(\mathbf{X}_1, \mathbf{X}_2) < \varepsilon$ then the trajectories of two phase-space points starting at $\mathbf{X}_1$ and $\mathbf{X}_2$ would remain within a distance of $\varepsilon$ over the time interval $-T \leq t \leq +T$.

We suppose that there exists an invariant set $\mathcal{R}$ of interest in the constant energy surface, i. e. a set such that

$$\Phi^t(\mathcal{R}) = \mathcal{R} \tag{38}$$

Consider now a time interval $(-T, +T)$ and a subset of the invariant set, $\mathcal{S} \subset \mathcal{R}$, composed of points which are separated by a $\rho_T$ distance larger than $\varepsilon$. That is, we construct a set of points $\mathcal{S} = \{\mathbf{Y}_1, \ldots, \mathbf{Y}_S\}$ of the invariant set $\mathcal{R}$ such that $\rho_T(\mathbf{Y}_i, \mathbf{Y}_j) > \varepsilon$ for $i, j = 1, \ldots, S$. If the invariant set is compact, i.e. bounded, then one can always find a subset $\mathcal{S}$ with a finite number of points. This set is called a $(\varepsilon, T)$-*separated subset* of the invariant set.

Since $(\varepsilon, T)$-separated subsets exist with a finite number of points, there exists at least one set, $\mathcal{S}_T(\varepsilon)$, with the maximum number of points, say $s_T(\varepsilon, \mathcal{R})$ points. Since the trajectories separate exponentially with $T$ we expect that $s_T(\varepsilon)$ will increase exponentially with $T$ also because the points may become separated by smaller and smaller distances according to Eq. (36). This rate of growth of $s_T(\varepsilon)$ is characterized by the *topological entropy* defined as

$$h_{\text{top}}(\Phi) = \lim_{\varepsilon \to 0} \limsup_{T \to \infty} \frac{1}{2T} \ln s_T(\varepsilon). \tag{39}$$

*2. Topological pressure and the dynamical invariant measures*

The idea of the thermodynamic formalism is to introduce a functional of physical observables which is the generating functional of the average and of the multitime correlation functions of the given observable $A(\mathbf{X})$. The functional is called the Ruelle topological pressure and is defined as

$$\mathcal{P}(A) = \lim_{\varepsilon \to 0} \limsup_{T \to \infty} \frac{1}{2T} \ln \mathcal{Z}_T(A, \varepsilon). \tag{40}$$

with the partition functional

$$\mathcal{Z}_T(A, \varepsilon) = \text{Sup}_{\mathcal{S}} \sum_{\mathbf{Y} \in \mathcal{S}} \exp \int_{-T}^{+T} A(\Phi^t \mathbf{Y}) dt \tag{41}$$

where $\mathcal{S}$ is a $(\varepsilon, T)$-separated subset of the invariant set $\mathcal{R}$. The topological pressure has remarkable properties. In particular, it is a convex functional of the observable, i. e. $\mathcal{P}[\nu A + (1 - \nu)B] \geq \nu \mathcal{P}(A) + (1 - \nu)\mathcal{P}(B)$ for $0 \leq \nu \leq 1$ and any two observables $A$ and $B$.

When the observable is everywhere vanishing, $A = 0$, the topological pressure reduces to the topological entropy because

$$s_T(\varepsilon) = \mathcal{Z}_T(A = 0, \varepsilon), \quad \text{so that} \quad h_{\text{top}}(\Phi) = \mathcal{P}(0) \tag{42}$$

Let us now consider another observable $B$ of the dynamical system. The average of this observable is defined by

$$\mu_A(B) = \frac{d}{d\nu}\mathcal{P}(A+\nu B)\Big|_{\nu=0} \qquad (43)$$

Introducing the definition of the pressure, we get that $\mu_A(B) = \int B(\mathbf{X})\mu_A(d\mathbf{X})$ with the measure

$$\mu_A(d\mathbf{X}) = \lim_{\varepsilon\to 0}\limsup_{T\to\infty}\text{Sup}_\mathcal{S}\sum_{\mathbf{Y}\in\mathcal{S}}\frac{\exp\int_{-T}^{+T}A(\Phi^t\mathbf{Y})dt}{\mathcal{Z}_T(A,\varepsilon)}\frac{1}{2T}\int_{-T}^{+T}\delta(\mathbf{X}-\Phi^t\mathbf{Y})dt\ d\mathbf{X} \qquad (44)$$

which is referred to as a dynamical measure. We observe that each trajectory of the $(\varepsilon,T)$-separated subset is weighted by a Boltzmann-type probability given by

$$\pi_A(\mathbf{Y},T,\varepsilon) = \frac{\exp\int_{-T}^{+T}A(\Phi^t\mathbf{Y})dt}{\mathcal{Z}_T(A,\varepsilon)} \qquad (45)$$

where $\int_{-T}^{+T}A(\Phi^t\mathbf{X})dt$ plays the role of $-\beta E$. Accordingly, such measures have been called Gibbs canonical measure. Here, the role of the energy is played by the average of the observable $A$ over the time interval $(-T,+T)$. We emphasize that the Gibbs measure, Eq. (44) is determined by a time interval $(-T,+T)$ rather than by a number of particles, interaction range, etc., as in equilibrium statistical mechanics. In the limit $T\to\infty$ the probability measures, Eq. (44), tend to measures which are invariant under time evolution, known as "equilibrium states" as proved by Bowen and Ruelle for Axiom-A hyperbolic systems [12]. In this limit a connection can be established between the measures Eq. (44) and the invariant probability measure on the invariant set, a connection which will be important for us subsequently.

In this formalism, correlation functions between two observables $B_1$ and $B_2$ can be obtained as second derivatives

$$\frac{\partial^2}{\partial\nu_1\partial\nu_2}\mathcal{P}(A+\nu_1 B_1+\nu_2 B_2)\Big|_{\nu_1=\nu_2=0} = \int_{-\infty}^{+\infty}\Big[\mu_A(B_1 B_2\circ\Phi^t)-\mu_A(B_1)\mu_A(B_2)\Big]dt \qquad (46)$$

in a straightforward notation, provided the various limits exist. Moreover, the Kolmogorov-Sinai entropy per unit time of the dynamical system with respect to the invariant measure $\mu_A$ is defined by [24,27]

$$h_{\text{KS}}(\mu_A) = \lim_{\varepsilon\to 0}\limsup_{T\to\infty}\left(-\frac{1}{2T}\right)\text{Sup}_\mathcal{S}\sum_{\mathbf{Y}\in\mathcal{S}}\pi_A(\mathbf{Y},T,\varepsilon)\ \ln\pi_A(\mathbf{Y},T,\varepsilon) \qquad (47)$$

We can now combine Eqs. (43) and (47) with $B=A$ to obtain the fundamental identity of importance to us,

$$h_{\text{KS}}(\mu_A) = -\mu_A(A) + \mathcal{P}(A) \qquad (48)$$

*3. Pressure functions based on the Lyapunov exponents*

By varying the observable $A$, we can obtain many invariant measures, and it is not yet clear which among these many possible measures is the one appropriate to a specific numerical simulation of the dynamical system starting from a given statistical ensemble. The answer to this question will be delayed till the next two sections where the question is answered for closed and open hyperbolic systems. Nevertheless, one thing that is common to the closed and open cases is the special role played by the sum over the local stretching rates multiplied by a parameter $-\beta$ in order to emphasize the formal analogy with the Gibbs states

$$A(\mathbf{X}) = -\beta u(\mathbf{X}) = -\beta\sum_{\lambda_i>0}\chi_i(\mathbf{X}) \qquad (49)$$

in terms of the local dispersion rate (32). This observable measures the dispersion of trajectories emanating from points in this region over a time interval $(-T,+T)$. The larger the dynamical instability on the trajectory, the smaller the probability to visit the neighborhood of this trajectory. This reasoning is at the basis of the choice (49). In this case, the pressure functional becomes the following pressure function

$$P(\beta) = \mathcal{P}\Big[-\beta\sum_{\lambda_i>0}\chi_i(\mathbf{X})\Big] \qquad (50)$$

which defines an invariant probability measure $\mu_\beta$ depending on the parameter $\beta$. The sum of averaged stretching rates is then given by

$$-\frac{dP}{d\beta}(\beta) = \mu_\beta\left(\sum_{\lambda_i>0} \chi_i\right) \tag{51}$$

Using the time invariance of the dynamical measure $\mu_\beta$ applied to Eq. (31), we obtain that the averages of the local stretching rates are identical with the averages of the corresponding local positive Lyapunov exponent

$$\mu_\beta(\chi_i) = \mu_\beta(\lambda_i) \tag{52}$$

Therefore, the fundamental identity (48) becomes here

$$h_{\text{KS}}(\mu_\beta) = \beta\,\mu_\beta\left(\sum_{\lambda_i>0}\lambda_i\right) + P(\beta) = \beta\,\mu_\beta(u) + P(\beta) \tag{53}$$

We notice that the parameter $\beta$ has nothing directly to do with the inverse of a thermodynamic temperature. It is worth noting, however, that the thermodynamic formalism does make possible deep connections between the calculation of thermodynamic functions and dynamical properties, a subject which is discussed at some length elsewhere [10–13,28–30].

The individual average Lyapunov exponents can be obtained by defining a multivariate pressure function $P(\beta_1,\ldots,\beta_L)$ from the observable

$$A(\mathbf{X}) = -\sum_{\lambda_i>0}\beta_i\chi_i(\mathbf{X}) \tag{54}$$

depending on the $L = Nf - 1$ parameters $\boldsymbol{\beta} = (\beta_1,\ldots,\beta_L)$. The averages Lyapunov exponents are now given by

$$-\frac{\partial P}{\partial \beta_i}(\boldsymbol{\beta}) = \mu_{\boldsymbol{\beta}}(\lambda_i) \tag{55}$$

When all the parameters are equal $\beta = \beta_1 = \ldots = \beta_L$, the pressure function (50) is recovered: $P(\beta) = P(\beta_1,\ldots,\beta_L)$ (see Fig. 2).

### 4. Entropy function and Legendre transform

It is convenient to introduce also a entropy function by considering the number of points of the $(\varepsilon,T)$-separated subset such that the time average of their associated local stretching rates, calculated over a time interval $(-T,+T)$, take values in the interval $(\varphi_i, \varphi_i + d\varphi_i)$ according to

$$\text{Number}\left\{\mathbf{Y}\in\mathcal{S}:\;\frac{1}{2T}\int_{-T}^{+T}\chi_i(\Phi^t\mathbf{Y})dt\,\in\,(\varphi_i,\varphi_i+d\varphi_i),\;i=1,\ldots,L\right\} = \rho(T,\mathcal{S},\boldsymbol{\varphi})\,\exp\left[2TS(\boldsymbol{\varphi})\right]\,d^L\varphi \tag{56}$$

in the limit where $\varepsilon \to 0$ and $T \to \infty$, where $\rho(T,\mathcal{S},\boldsymbol{\varphi})$ is a slowly varying function of the time $T$. The entropy function $S(\boldsymbol{\varphi})$ is known to be a concave function [6,31,32].

The entropy function is related to the pressure function $P(\boldsymbol{\beta})$ by a Legendre transform. Indeed, the sum over all the points of the $(\varepsilon, T)$-separated subset $\mathcal{S}$ can be replaced by an integral over $\boldsymbol{\varphi}$

$$P(\boldsymbol{\beta}) = \lim_{T\to\infty}\frac{1}{2T}\ln\int d^L\varphi\,\rho(T,\mathcal{S},\boldsymbol{\varphi})\,\exp[2TS(\boldsymbol{\varphi})]\,\exp(-2T\boldsymbol{\varphi}\cdot\boldsymbol{\beta}) \tag{57}$$

In the limit $T \to \infty$, the integral can be evaluated by the steepest-descent method which selects the maximum $\boldsymbol{\varphi}_{\boldsymbol{\beta}}$ of the function in the argument of the exponential as solution of

$$\frac{\partial S}{\partial \boldsymbol{\varphi}}(\boldsymbol{\varphi}_{\boldsymbol{\beta}}) = \boldsymbol{\beta} \tag{58}$$

Therefore, the pressure function is given by

$$P(\boldsymbol{\beta}) = S(\boldsymbol{\varphi}_{\boldsymbol{\beta}}) - \boldsymbol{\varphi}_{\boldsymbol{\beta}} \cdot \boldsymbol{\beta} \tag{59}$$

Reciprocally, once the pressure function is known, the entropy function is obtained as

$$S(\boldsymbol{\varphi}) = P(\boldsymbol{\beta}_{\boldsymbol{\varphi}}) + \boldsymbol{\varphi} \cdot \boldsymbol{\beta}_{\boldsymbol{\varphi}} \qquad \text{with} \qquad \boldsymbol{\varphi} = -\frac{\partial P}{\partial \boldsymbol{\beta}}(\boldsymbol{\beta}_{\boldsymbol{\varphi}}) \tag{60}$$

In particular, the topological entropy is given by

$$h_{\text{top}}(\Phi) = P(0) = S(\boldsymbol{\varphi}_0) \qquad \text{with} \qquad \boldsymbol{\varphi}_0 = -\frac{\partial P}{\partial \boldsymbol{\beta}}(0) \qquad \text{or} \qquad \frac{\partial S}{\partial \boldsymbol{\varphi}}(\boldsymbol{\varphi}_0) = 0 \tag{61}$$

and the KS entropy by

$$h_{\text{KS}}(\mu_{\boldsymbol{\beta}}) = S(\boldsymbol{\varphi}_{\boldsymbol{\beta}}) \qquad \text{since} \qquad \boldsymbol{\varphi}_{\boldsymbol{\beta}} = \mu_{\boldsymbol{\beta}}(\boldsymbol{\lambda}) \tag{62}$$

which justifies the name. Here also, we can define a univariate entropy function $S(\varphi) = S(\varphi, \ldots, \varphi)$, which is related to the univariate pressure function by a Legendre transform.

We need a proper interpretation of the different terms appearing in the fundamental identity (53), especially of the pressure $P(\beta)$ and we need to fix the value of the parameter $\beta$.

### D. Closed Hyperbolic Systems

#### 1. The microcanonical measure as a Sinai-Bowen-Ruelle dynamical measure

In a closed, conservative, ergodic system the natural measure is the microcanonical measure $\mu$. The appropriate invariant measure on this surface is, of course, the microcanonical measure $\mu_{\text{eq}}(d\mathbf{X})$ given by

$$\mu_{\text{eq}}(d\mathbf{X}) = \frac{d\sigma(\mathbf{X})}{|\text{grad}H|} \tag{63}$$

where $d\sigma(\mathbf{X})$ is a surface area element about the point $\mathbf{X}$ and $H$ is the Hamiltonian of the system. The gradient is a $2Nf$ dimensional gradient evaluated on the surface $H = E$. Our purpose is to identify the microcanonical measure with one of the dynamical measures introduced in subsection II.B. With this purpose, we consider one of the small regions surrounding a point $\mathbf{Y}$ in a finite $(\varepsilon, T)$-separated subset $\mathcal{S}$ and the probability of this region can be computed using Eq. (63) once the region is identified. In subsection II.B, we considered that this same region is a domain on the constant energy surface in which the trajectories of all points will remain separated by a distance less than $\varepsilon$ over a time interval $(-T, +T)$, which is known as a ball $\mathcal{B}_T(\mathbf{Y}, \varepsilon)$. The microcanonical probability of such a ball can be estimated using the results of subsection II.A as follows

$$\begin{aligned}
\mu_{\text{eq}}\Big[\mathcal{B}_T(\mathbf{Y}, \varepsilon)\Big] &= \mu_{\text{eq}}\Big\{\mathbf{X} \in \mathcal{M}: \quad \|\Phi^t\mathbf{X} - \Phi^t\mathbf{Y}\| \leq \varepsilon, \; \forall t \in [-T, +T]\Big\} \\
&\simeq \mu_{\text{eq}}\Big\{\mathbf{X} \in \mathcal{M}: \quad \|\mathbf{M}(t, \mathbf{Y}) \cdot (\mathbf{X} - \mathbf{Y})\| \leq \varepsilon, \; \forall t \in [-T, +T]\Big\} \\
&= \mu_{\text{eq}}\Big\{\mathbf{X} \in \mathcal{M}: \quad \sum_i \sigma_i(t, \mathbf{Y}) |\mathbf{u}_i(t, \mathbf{Y}) \cdot (\mathbf{X} - \mathbf{Y})|^2 \leq \varepsilon^2, \; \forall t \in [-T, +T]\Big\} \\
&\sim \prod_{\sigma_i(+T, \mathbf{Y}) > 1} \frac{1}{\sqrt{\sigma_i(+T, \mathbf{Y})}} \prod_{\sigma_i(-T, \mathbf{Y}) > 1} \frac{1}{\sqrt{\sigma_i(-T, \mathbf{Y})}} \\
&\sim \exp - \int_{-T}^{+T} \sum_{\lambda_i > 0} \chi_i(\Phi^t\mathbf{Y}) dt \\
&\sim \exp - \int_{-T}^{+T} u(\Phi^t\mathbf{Y}) dt \\
&\sim \pi_u(\mathbf{Y}, T, \varepsilon)
\end{aligned} \tag{64}$$

At the first line, the definition (37) of the distance has been used. We supposed that $\varepsilon$ is small enough and we used Eq. (25) to get the second line. The third line results from the spectral decomposition (26). The fourth line is based on the facts that the quadratic form defines a small ellipsoid with axes determined by the quantities $\sigma_i(t, \mathbf{Y})$. Half of these quantities increase exponentially for $t > 0$ and the other half increase exponentially for $t < 0$, while the vectors $\mathbf{u}_i(t, \mathbf{Y})$ are slowly varying functions of time. The fifth line is a consequence of the relation (27) of the eigenvalues $\sigma_i(t, \mathbf{Y})$ to the local stretching rates and of the pairing rule that, to every stretching rate, there corresponds a contracting rate with the same absolute value by time reversibility. Finally, the last two lines result from the definition (49) of the observable $u(\mathbf{X})$ and of the probability (45).

To get this last result, we used the fact that the normalization factor $\mathcal{Z}_T(u,\varepsilon)$ is a slowly varying (subexponential) function of time in the case of closed hyperbolic systems. Indeed, the sum of all the probabilities (64) is approximately constant. On the other hand, all the equalities in Eq. (64) hold up to factors which are slowly varying with time. According to the fifth line, the sum of the exponential factors involving the local stretching rates is slowly varying. With the last line, this implies that the dynamical partition function $\mathcal{Z}_T(u,\varepsilon)$ is also slowly varying with time. Therefore, all of the exponential dependence should be completely taken into account by the dispersion factor given on the right-hand side of Eq. (64). This observation suggests that the member of the family of measures, Eq. (44) based on the observable $A = -u$ which corresponds to the value $\beta = 1$ is the natural invariant measure corresponding to the microcanonical measure

$$\mu_{\text{eq}} = \mu_{\beta=1} \tag{65}$$

This result is general and extends to open hyperbolic systems.

We now develop some consequences of this identification. Since the system is closed and the equilibrium measure of the constant energy surface is finite, both the microcanonical and the dynamic measures must be normalizable to unity, say. As a consequence, we already concluded that the normalization factor $\mathcal{Z}_T(u,\varepsilon)$ does not depend on time in an exponential way[3], but increases in a subexponential way with $T$ when $\beta = 1$, and, as a result the pressure at this value of $\beta$ must be zero,

$$P(1) = 0 \quad \text{for closed systems} \tag{66}$$

Inserting this result in the fundamental identity, Eq. (53) with $\beta = 1$ we recover Pesin's identity [5,33]

$$h_{\text{KS}}(\mu_{\text{eq}}) = \sum_{\lambda_i > 0} \mu_{\text{eq}}(\lambda_i) \tag{67}$$

so that the KS entropy of the invariant natural measure is the sum of the positive Lyapunov exponents averaged over the same measure. We remark that the dynamical invariant measure which is the microcanonical measure is absolutely continuous with respect to the Lebesgue measure along the unstable manifolds. When this property holds which is a corollary of the Pesin identity, the dynamical measure is referred to as a Sinai-Bowen-Ruelle (SRB) measure [5,30].

### 2. The pressure function for closed systems

Another remarkable and useful identity for the pressure $P(\beta)$ can be obtained in an alternative way as an average over the microcanonical measure (63) according to

$$P(\beta) = \limsup_{T \to \infty} \frac{1}{2T} \ln \int_{\mathcal{M}} \mu_{\text{eq}}(d\mathbf{X}) \exp\left[(1-\beta) \int_{-T}^{+T} \sum_{\lambda_i > 0} \chi_i(\Phi^t \mathbf{X}) dt\right] \tag{68}$$

The original definition of the pressure can be recovered as follows. Considering an $(\varepsilon, T)$-separated subset $\mathcal{S}$ of the phase space $\mathcal{M}$, the integral can be discretized into a sum over the points $\{\mathbf{Y}\}$ of $\mathcal{S}$ replacing the volume elements $d\mathbf{X}$ by small balls $\mathcal{B}_T(\mathbf{Y},\varepsilon)$. This sum would be equal to the integral after the appropriate limits are taken. Therefore, the right-hand side of Eq. (68) is given by

$$\lim_{\varepsilon \to 0} \limsup_{T \to \infty} \frac{1}{2T} \ln \text{Sup}_{\mathcal{S}} \sum_{\mathbf{Y} \in \mathcal{S}} \mu_{\text{eq}}[\mathcal{B}_T(\mathbf{Y},\varepsilon)] \exp\left[(1-\beta) \int_{-T}^{+T} u(\Phi^t \mathbf{Y}) dt\right] \tag{69}$$

Now, according to Eq. (64), the probabilities of the small balls are exponentially growing like $\exp - \int_{-T}^{+T} u(\Phi^t \mathbf{Y}) dt$, which introduce an extra inverse power of the dispersing factor and explains the power $(1-\beta)$ in Eq. (68). Finally, we obtain the definition of the pressure, namely

---

[3]As it would if particles were escaping from the system at an exponential rate. See subsection II.D.

$$P(\beta) = \lim_{\varepsilon \to 0} \limsup_{T \to \infty} \frac{1}{2T} \ln \operatorname{Sup}_{\mathcal{S}} \sum_{\mathbf{Y} \in \mathcal{S}} \exp\left[-\beta \int_{-T}^{+T} u(\Phi^t \mathbf{Y}) dt\right] \qquad (70)$$

We note that for the purpose of numerical calculations, we can substitute the stretching factors $\sigma_i(t, \mathbf{X})$ for the exponentials of the integrated local stretching factors and use

$$P(\beta) = \lim_{T \to \infty} \frac{1}{T} \ln \int_{\mathcal{M}} \mu_{\mathrm{eq}}(d\mathbf{X}) \prod_{\sigma_i > 1} \sqrt{\sigma_i(T, \mathbf{X})}^{1-\beta} \qquad (71)$$

instead of Eq. (68). Here, there is no factor of two in the denominator because the stretching factors correspond to the time interval $(0, +T)$ rather than $(-T, +T)$ as before.

Equations (68) or (71) provide a convenient way to compute the pressure function as a phase-space average of an expression involving the local stretching factors. A typical pressure function is illustrated in Fig. 3a for a closed hyperbolic system. Note that it is a convex, monotonic function of $\beta$ and vanishes at $\beta = 1$. This is not true of open systems. The KS entropy is obtained by finding the slope of the pressure at $\beta = 1$, and carrying out the linear extrapolation illustrated in Fig. 3a. Finally, we note that the value of $P(\beta)$ at $\beta = 0$, is the topological pressure $h_{\mathrm{top}}(\Phi)$ of the dynamical system, as can be seen from Eqs. (39), and (61). The fact that $h_{\mathrm{KS}} \leq h_{\mathrm{top}}$ puts a bound on the slope of $P(\beta)$ at $\beta = 1$

### 3. Generating functions of transport coefficients

In paper I, we showed that each transport or rate coefficient $\alpha$ is associated to a diffusive-type motion for a corresponding Helfand moment $G_t^{(\alpha)}$. Within the large-deviation formalism, it is possible to characterize in detail the random time evolution of these Helfand moments by introducing, the generating function

$$Q(\kappa, \beta) = \lim_{T \to \infty} \frac{1}{2T} \ln \mu_\beta \left\{ \exp\left[\kappa (G_{+T}^{(\alpha)} - G_{-T}^{(\alpha)})\right] \right\} \qquad (72)$$

In particular, if there is no drift of the moment, i. e., if $\partial_\kappa Q\big|_{\kappa=0} = 0$, the transport coefficients with respect to the SRB measure $\mu_\beta$ is given by

$$\alpha_\beta = \frac{1}{2} \frac{\partial^2 Q}{\partial \kappa^2}\bigg|_{\kappa=0} \qquad (73)$$

The standard transport coefficient is given by the value at $\beta = 1$ at which we recover the microcanonical measure. The higher derivatives give the higher-order transport coefficients like the Burnett coefficients [34].

We can also define a large-deviation function according to

$$\mu_\beta \left\{ \mathbf{Y} \in \mathcal{S} : \frac{G_{+T}^{(\alpha)} - G_{-T}^{(\alpha)}}{2T} \in (\eta, \eta + d\eta) \right\} = \rho(T, \mathcal{S}, \eta, \beta) \, \exp\left[2TH(\eta, \beta)\right] \, d\eta \qquad (74)$$

in the limit $T \to \infty$. In Eq. (74), $\rho$ is a slowly varying function and the Helfand moment is calculated for the trajectory $\mathbf{Y}$ of an $(\varepsilon, T)$-separated subset of the phase space $\mathcal{M}$. The function $H(\eta, \beta)$ is related to the generating function by a Legendre transform according to

$$Q(\kappa, \beta) = H(\eta, \beta) + \kappa \eta, \quad \text{for} \quad \kappa = -\frac{\partial H}{\partial \eta} \quad \text{and} \quad \eta = \frac{\partial Q}{\partial \kappa} \qquad (75)$$

These considerations are particularly useful in the study of anomalous transport [35].

### E. Open Hyperbolic Systems

#### 1. Escape-time functions and the repeller

The main focus of our analysis in Paper I was on the chaotic behavior of open systems, i.e. systems with trajectories which escape from a bounded region of phase space in a finite amount of time. The behavior of such systems can be easily visualized by considering an escape-time function as shown in Fig. 4 for a simple model of a logistic map with escape from the interval $0 \leq x \leq 1$. The connection between transport coefficients and dynamical quantities is based on the escape-rate formula, Eq. (3), and here we describe the derivation as a special case of the fundamental identity, Eq. (48).

An open system is defined by a trajectory dynamics taking place in a phase-space region $\mathcal{B}$, with a boundary $\partial \mathcal{B}$. We consider initial conditions $\mathbf{X}_0$, within $\mathcal{B}$, and define an escape time $T_\mathcal{B}^{(+)}(\mathbf{X}_0)$ as the first time when the trajectory crosses the boundary

$$\Phi^t \mathbf{X}_0 \in \mathcal{B} \text{ for } 0 < t < T_\mathcal{B}^{(+)}(\mathbf{X}_0) \tag{76}$$
$$\Phi^{T_\mathcal{B}^{(+)}(\mathbf{X}_0)} \mathbf{X}_0 \in \partial \mathcal{B}, \text{ and}$$
$$\Phi^{T_\mathcal{B}^{(+)}(\mathbf{X}_0)+\delta} \mathbf{X}_0 \notin \partial \mathcal{B}, \mathcal{B}$$

for arbitrarily small $\delta > 0$. Since we are considering time-reversible, Hamiltonian systems, we may also define a "negative" escape time for an initial condition $\mathbf{X}_0 \in \mathcal{B}$ in an analogous way to that in Eq. (76) by

$$\Phi^t(\mathbf{X}_0) \in \mathcal{B} \text{ for } T_\mathcal{B}^{(-)}(\mathbf{X}_0) < t < 0 \tag{77}$$
$$\Phi^{T_\mathcal{B}^{(-)}(\mathbf{X}_0)}(\mathbf{X}_0) \in \partial B, \text{ and}$$
$$\Phi^{T_\mathcal{B}^{(-)}(\mathbf{X}_0)+\delta}(\mathbf{X}_0) \notin \partial \mathcal{B}, \mathcal{B}$$

for arbitrarily small $\delta < 0$. The escape time is a highly singular function of the initial conditions if there exist periodic and non-periodic trajectories which are forever trapped inside the domain $\mathcal{B}$ such that $\Phi^t \mathbf{X}_0 \in \mathcal{B}$ for $-\infty < t < \infty$. Since most of the trajectories are expected to escape from $\mathcal{B}$ for hyperbolic systems, these trapped set of trajectories form a set of zero Lebesgue measure. This set – called the repeller – may contain a subset of chaotic trajectories in which case the repeller is a fractal set. Because of hyperbolicity, the points on the trapped trajectories must have stable and unstable manifolds. All of the initial conditions in $\mathcal{B}$ belonging to the stable manifolds of the trapped orbits remain in $\mathcal{B}$ for all positive times. For these initial conditions the escape time is infinite. However these initial points form a set of zero Lebesgue measure so that the escape-time function takes finite values for almost all of the points in $\mathcal{B}$ (see Fig. 4).

We now consider the set of initial conditions for which the escape time is larger than a predetermined time $T > 0$,

$$\Upsilon_\mathcal{B}^{(+)}(T) = \left\{ \mathbf{X}_0 \in \mathcal{B} : T < T_\mathcal{B}^{(+)}(\mathbf{X}_0) \right\}. \tag{78}$$

This set contains all of the members of the statistical ensemble which are still inside the domain $\mathcal{B}$ at the time $T$, i.e.

$$\Upsilon_\mathcal{B}^{(+)}(T) = \bigcap_{0<t<T} \Phi^{-t}(\mathcal{B}). \tag{79}$$

In the long time limit, this set contains the trapped trajectories of the fractal repeller $\mathcal{R}_\mathcal{B}$ and their stable manifolds, restricted to the domain $\mathcal{B}$:

$$\lim_{T\to\infty} \Upsilon_\mathcal{B}^{(+)}(T) = \text{Cl}[W_s(\mathcal{R}_\mathcal{B})] \bigcap \mathcal{B} \tag{80}$$

where Cl[·] denotes the closure of the set. As a consequence, the set $\Upsilon_\mathcal{B}^{(+)}(T)$ undergoes a fragmentation into smaller and smaller sets as time increases to end up as the fractal set (80). In a similar way, we can define a set of initial conditions such that the time reversed motion has a "negative" escape time which satisfies $|T_\mathcal{B}^{(-)}(\mathbf{X}_0)| > T$

$$\Upsilon_\mathcal{B}^{(-)}(T) = \left\{ \mathbf{X}_0 \in \mathcal{B} : |T_\mathcal{B}^{(-)}(\mathbf{X}_0)| > T, T_\mathcal{B}^{(-)} < 0 \right\} = \bigcap_{-T<t<0} \Phi^{-t}(\mathcal{B}) \tag{81}$$

In analogy with Eq. (80), we have that

$$\lim_{T\to\infty} \Upsilon_\mathcal{B}^{(-)}(T) = \text{Cl}[W_u(\mathcal{R}_\mathcal{B})] \bigcap \mathcal{B} \tag{82}$$

The intersection of the two sets,

$$\Upsilon_{\mathcal{B}}(T) = \Upsilon_{\mathcal{B}}^{(-)}(T) \cap \Upsilon_{\mathcal{B}}^{(+)}(T) = \bigcap_{-T<t<+T} \Phi^{-t}(\mathcal{B}) \tag{83}$$

contains all of the members of the statistical ensemble which are inside the region $\mathcal{B}$ over the time interval $-T < t < +T$. The repeller is defined as the set of points $\Upsilon_{\mathcal{B}}(T)$ as $T \to \infty$

$$\lim_{T \to \infty} \Upsilon_{\mathcal{B}}(T) = \text{Cl}[\mathcal{R}_{\mathcal{B}}] \bigcap \mathcal{B} \tag{84}$$

It is instructive to consider this construction for simple systems such as a two-dimensional Smale horseshoe map of the unit square as discussed by Lanford, Tel, and others [25,36,37]. One sees that for this map the sets $\Upsilon_{\mathcal{B}}^{(+)}(T)$ and $\Upsilon_{\mathcal{B}}^{(-)}(T)$ are thin strips parallel to the stable and unstable directions, respectively, each becoming the product of a Cantor set with a one-dimensional interval as $T \to \infty$. The repeller is the intersection of these two sets, and is a Cantor set which can be coded as a bi-infinite sequence of zeros and ones. The dynamics on the repeller is then isomorphic to the left-Bernoulli shift on these bi-infinite sequences.

*2. The nonequilibrium invariant measure of the repeller*

We now define the invariant probability measure of the repeller. We first consider a probability measure $\nu_0(d\mathbf{X})$ on the region $\mathcal{B}$ corresponding to a particular statistical ensemble of $N_0$ initial conditions at phase-space points $\{\mathbf{X}^{(j)}\}$ where these points are distributed uniformly in $\mathcal{B}$ with respect to the microcanonical ensemble, say. This measure $\nu_0$ is taken to be of the form

$$\nu_0(d\mathbf{X}) = \lim_{N_0 \to \infty} \frac{1}{N_0} \sum_{j=1}^{N_0} \delta\left(\mathbf{X} - \mathbf{X}^{(j)}\right) d\mathbf{X} \tag{85}$$

Of these $N_0$ initial points, the number $N_T$ still contained in $B$ over the time interval $(0, +T)$ decays according to

$$\lim_{N_0 \to \infty} \frac{N_T}{N_0} = \nu_0[\Upsilon_{\mathcal{B}}^{(+)}(T)] = \int_{\Upsilon_{\mathcal{B}}^{(+)}(T)} \nu_0(d\mathbf{X}). \tag{86}$$

We note that the limit $N_0 \to \infty$ is essential to define a smooth function of time $T$, since $N_T$ typically shows large statistical fluctuations when $N_T < 10$, and drops to zero after a finite time. Similar expressions hold for the time intervals $(-T, 0)$ and $(-T, +T)$.

Assuming that almost all the trajectories escape, i. e. that $\lim_{t \to \infty} \nu_0[\Upsilon_{\mathcal{B}}^{(+)}(T)] = 0$, the decay curve, Eq. (86), may be exponential or slower than exponential in general systems. However, in hyperbolic systems (where all orbits are of saddle type) the decay is exponential. Thus we can define an escape rate according to

$$\gamma = -\lim_{T \to \infty} \frac{1}{2T} \ln \nu_0 \left[\Upsilon_{\mathcal{B}}^{(+)}(T) \cap \Upsilon_{\mathcal{B}}^{(-)}(T)\right] \tag{87}$$

$$= -\lim_{T \to \infty} \frac{1}{T} \ln \nu_0 \left[\Upsilon_{\mathcal{B}}^{(+)}(T)\right] \tag{88}$$

$$= -\lim_{T \to \infty} \frac{1}{T} \ln \nu_0 \left[\Upsilon_{\mathcal{B}}^{(-)}(T)\right] \tag{89}$$

All these definitions are equivalent because of time reversibility. For noninvertible systems like 1D maps, only Eq. (88) is of application.

In the long-time limit, the trajectories remaining in the domain $\mathcal{B}$ are distributed according to a probability measure which is invariant for the dynamics on the repeller. In order to construct such an invariant probability measure on the repeller, we use ideas familiar from ergodic theory. In the usual arguments of ergodic theory one considers the time average of some dynamical quantity for a system whose phase-space point is confined to a constant energy surface.

If the system is ergodic then the long time average of any dynamical quantity is equal to its ensemble average taken with respect to the microcanonical measure. Consider now a system whose phase-space trajectory is confined to the repeller. If the trajectories on the repeller are ergodic with respect to a natural measure on the repeller, then the long time average of any dynamical quantity on the repeller should be equal to the ensemble average of this quantity with respect to the natural measure. To construct the natural nonequilibrium measure on the repeller, then, we begin with the definition of the time average of some observable $A(\mathbf{X})$ on the repeller as

$$\mu_{\rm ne}(A) = \lim_{T\to\infty} \lim_{N_T\to\infty} \frac{1}{N_T} \sum_{j=1}^{N_T} \frac{1}{2T} \int_{-T}^{+T} A(\Phi^t \mathbf{X}^{(j)})\, dt. \tag{90}$$

In Eq. (90), the sum extends over the $N_T$ phase-space points whose trajectories remain in region $\mathcal{B}$ over the time interval $-T < t < +T$. The time average can be rewritten in terms of the initial measure as

$$\mu_{\rm ne}(A) = \int A(\mathbf{X}) \mu_{\rm ne}(d\mathbf{X}) \tag{91}$$

with

$$\mu_{\rm ne}(d\mathbf{X}) = \lim_{T\to\infty} \frac{1}{\nu_0[\Upsilon_\mathcal{B}(T)]} \int \nu_0(d\mathbf{Z})\, I_{\Upsilon_\mathcal{B}(T)}(\mathbf{Z})\, \frac{1}{2T} \int_{-T}^{+T} \delta(\mathbf{X} - \Phi^t \mathbf{Z}) dt \tag{92}$$

Here $I_\Upsilon(\mathbf{X})$ is the characteristic function of the set $\Upsilon$ in phase space, i.e., $I_\Upsilon(\mathbf{X}) = 1$ if $\mathbf{X} \in \Upsilon$ and $= 0$, otherwise. In this way, we have defined a normalized probability measure on the repeller $\mu_{\rm ne}$. This is the desired invariant measure with the fractal repeller as its support. We also note that the subscript 'ne' of $\mu_{\rm ne}$ refers to the nonequilibrium character of the natural invariant measure. This measure is the natural generalization of the microcanonical canonical ensemble measure to open systems where the dynamics takes place on the fractal repeller, $\mathcal{R}_\mathcal{B}$. Using this measure, we can consider the long time limit and define the average of an observable according to Eq. (91). This procedure amounts to performing the statistics on the set of $N_T$ initial conditions which are still in the domain $\mathcal{B}$ at time $T$. As $T \to \infty$ these trajectories approach more and more closely to the trajectories on the repeller. As a consequence, $\mu_{\rm ne}$ is an invariant probability measure on the repeller. We will return to this invariant measure presently.

*3. Connection with the dynamical invariant measure and the pressure function*

Now, we wish to construct a dynamical measure on the repeller similar to that used for closed systems in subsection II.C. This will enable us to continue the development of the previous sections so as to the fundamental identity, Eq. (48) to open systems. The thermodynamic formalism suggests to consider the family of dynamical invariant measures defined in analogy with Eq. (44)

$$\mu_\beta(d\mathbf{X}) = \lim_{\varepsilon\to 0} \limsup_{T\to\infty} \mathrm{Sup}_\mathcal{S} \sum_{\mathbf{Y}\in\mathcal{S}} \frac{\exp -\beta \int_{-T}^{+T} u(\Phi^t \mathbf{Y}) dt}{Z_T(\beta,\varepsilon)}\, \frac{1}{2T} \int_{-T}^{+T} \delta(\mathbf{X} - \Phi^t \mathbf{Y}) dt\, d\mathbf{X} \tag{93}$$

with the dynamical partition function

$$Z_T(\beta,\varepsilon) = \mathrm{Sup}_\mathcal{S} \sum_{\mathbf{Y}\in\mathcal{S}} \exp -\beta \int_{-T}^{+T} u(\Phi^t \mathbf{Y}) dt \tag{94}$$

in terms of the dispersion rate $u = \sum_{\lambda_i > 0} \chi_i$. Let us emphasize that the local stretching rates as well as the local Lyapunov exponents are well defined for the trajectories of any subset $\mathcal{S}$ of the repeller $\mathcal{R}_\mathcal{B}$ since those trajectories remain in the compact domain $\mathcal{B}$ forever. A disadvantage of this definition is that the repeller $\mathcal{R}$ needs to be known since a subset $\mathcal{S}$ of it is considered. Physically and numerically, the repeller of systems of scattering type appears out of the dynamics after a very long time as above described. The previously defined invariant measure avoids this *a priori* knowledge of the repeller since the invariant measure (92) is automatically constructed by the time evolution

We now show that both measures are equivalent for open hyperbolic systems when $\beta = 1$. We do this by using the neighborhoods of the points of a $(\varepsilon, T)$-separated subset, $\mathcal{S}$, of the repeller $\mathcal{R}$. Note that we take the time $T$ to be the same in both sets so that we can closely approximate the repeller by means of the separated subsets as we take the limit $T \to \infty$. In this situation, we can transform the expression (92) into

$$\mu_{\rm ne}(d\mathbf{X}) = \lim_{\varepsilon\to 0} \limsup_{T\to\infty} \mathrm{Sup}_\mathcal{S} \sum_{\mathbf{Y}\in\mathcal{S}} \frac{\nu_0[\mathcal{B}_T(\mathbf{Y},\varepsilon) \cap \Upsilon_\mathcal{B}(T)]}{\nu_0[\Upsilon_\mathcal{B}(T)]} \frac{1}{2T} \int_{-T}^{+T} \delta(\mathbf{X} - \Phi^t \mathbf{Y}) dt \tag{95}$$

It is clear that this measure is normalized. Since the points $\mathbf{Y}$ belongs to a $(\varepsilon, T)$-separated subset of the repeller $\mathcal{R}_\mathcal{B}$, they belong to the repeller itself. If the points $\mathbf{Y}$ are not close to the boundary $\partial\mathcal{B}$ and if $\varepsilon$ is small enough, the balls $\mathcal{B}_T(\mathbf{Y}, \varepsilon)$ are contained inside $\Upsilon_\mathcal{B}(T)$ so that

$$\nu_0[\mathcal{B}_T(\mathbf{Y},\varepsilon) \cap \Upsilon_\mathcal{B}(T)] \simeq \nu_0[\mathcal{B}_T(\mathbf{Y},\varepsilon)] \sim \exp{-\int_{-T}^{+T} u(\Phi^t \mathbf{Y})dt} \tag{96}$$

where considerations similar to the ones of Eq. (64) have been here applied to $\nu_0$. As a consequence, Eq. (95) becomes

$$\mu_{\text{ne}}(d\mathbf{X}) = \lim_{\varepsilon \to 0} \limsup_{T \to \infty} \text{Sup}_\mathcal{S} \sum_{\mathbf{Y} \in \mathcal{S}} \frac{\exp{-\int_{-T}^{+T} u(\Phi^t \mathbf{Y})dt}}{Z_T(1,\varepsilon)} \frac{1}{2T} \int_{-T}^{+T} \delta(\mathbf{X} - \Phi^t \mathbf{Y})dt = \mu_1(d\mathbf{X}) \tag{97}$$

Accordingly, we recover the dynamical invariant measure corresponding to the value $\beta = 1$ as in the case of closed hyperbolic systems [see Eq. (65)].

We note that the normalization factor is here required for the following reason. In order to estimate the normalization factor, we decompose the set $\Upsilon_\mathcal{B}(T)$ into small balls $\mathcal{B}_T(\mathbf{Y},\varepsilon)$ centered on the points $\mathbf{Y}$ of an $(\varepsilon, T)$-separated subset $\mathcal{S}$ as

$$\nu_0[\Upsilon_\mathcal{B}(T)] \sim \text{Sup}_\mathcal{S} \sum_{\mathbf{Y} \in \mathcal{S}} \nu_0[\mathcal{B}_T(\mathbf{Y},\varepsilon) \cap \Upsilon_\mathcal{B}(T)]$$

$$\sim \text{Sup}_\mathcal{S} \sum_{\mathbf{Y} \in \mathcal{S}} \exp{-\int_{-T}^{+T} u(\Phi^t \mathbf{Y})dt} = Z_T(1,\varepsilon) \sim \exp{2TP(1)} \tag{98}$$

Comparing with the definition Eq. (87) of the escape rate $\gamma$, this result shows that the normalization factor decays exponentially like $Z_T(1,\varepsilon) \sim \exp(-2T\gamma)$ and, moreover, that the pressure function $P(\beta)$ has, for an open system of hyperbolic type, the value

$$P(1) = -\gamma. \tag{99}$$

Moreover, we can now define the important quantities of the thermodynamic formalism of subsection II.A and the fundamental identity (53) follows once again. We can now identify all of the quantities appearing in this equation at $\beta = 1$. For this case we find

$$\gamma = \sum_{\lambda_i > 0} \mu_{\text{ne}}(\lambda_i) - h_{\text{KS}}(\mu_{\text{ne}}) \ . \tag{100}$$

Eq. (100) is the escape-rate formula for an open system, giving the relation between the escape rate, $\gamma$, and the sum of the positive Lyapunov exponents and Kolmogorov-Sinai entropy for trajectories on the fractal repeller, $\mathcal{R}_\mathcal{B}$, using the natural measure $\mu_1 = \mu_{\text{ne}}$. We remark that we used the notations $\lambda_i(\mathcal{R}_\mathcal{B}) = \mu_{\text{ne}}(\lambda_i) = \mu_1(\lambda_i)$ and $h_{\text{KS}}(\mathcal{R}_\mathcal{B}) = h_{\text{KS}}(\mu_{\text{ne}}) = h_{\text{KS}}(\mu_1)$ in Eq. (2). This result generalizes Pesin's formula to open hyperbolic systems. Indeed, when the system is closed, the escape rate vanishes $\gamma = 0$ and Pesin's formula is recovered as well as Eq. (66). Here, the average of the sum of positive Lyapunov exponents over the nonequilibrium invariant measure can be calculated as before as a derivative of the pressure function

$$\sum_{\lambda_i > 0} \mu_{\text{ne}}(\lambda_i) = -\frac{dP(\beta)}{d\beta}\bigg|_{\beta=1} \tag{101}$$

Let us remark that the non-equilibrium invariant measure is not absolutely continuous with respect to the Lebesgue measure along the unstable manifolds but singular with a fractal for support. In this way, the nonequilibrium invariant measure differs from the equilibrium one.

Let us now describe a pratical way to calculate the pressure function in open systems. As we have seen, the pressure function require the knowldege of the local Lyapunov exponents. In practice, we only know the stretching factors $\sigma_i(T, \mathbf{X}^{(j)})$ of the trajectories of the initial ensemble (85). In analogy with Eq. (71), we propose here

$$P(\beta) = \lim_{T \to \infty} \frac{1}{T} \ln \int_{\Upsilon_\mathcal{B}^{(+)}(T)} \prod_{\sigma_i > 1} \sqrt{\sigma_i(T,\mathbf{X})}^{1-\beta} \nu_0(d\mathbf{X}) \tag{102}$$

This definition is equivalent to the previous one for the following reasons. By time reversibility, we can convert the average over the forward set $\Upsilon_\mathcal{B}^{(+)}(T)$ into an average over the set $\Upsilon_\mathcal{B}(T)$ by considering the time interval $(-T, +T)$ rather than $(0, +T)$. Using an $(\varepsilon, T)$-separated subset $\mathcal{S}$ to dicretize the integral, the right-hand member of (102) becomes

$$\lim_{\varepsilon \to 0} \limsup_{T \to \infty} \frac{1}{2T} \ln \operatorname{Sup}_{\mathcal{S}} \sum_{\mathbf{Y} \in \mathcal{S}} \nu_0[\mathcal{B}_T(\mathbf{Y}, \varepsilon) \cap \Upsilon_{\mathcal{B}}(T)] \exp(1-\beta) \int_{-T}^{+T} u(\Phi^t \mathbf{Y}) dt \qquad (103)$$

where we used the estimation (33). According to Eq. (96), we recover the original definition of the pressure

$$P(\beta) = \lim_{\varepsilon \to 0} \limsup_{T \to \infty} \frac{1}{2T} \ln Z_T(\beta, \varepsilon) \qquad (104)$$

with the partition function (94).

### F. Nonhyperbolic Systems

In many cases the system possesses periodic orbits with vanishing, but non trivial Lyapunov exponents. This is the case, in particular, for the stadium billiard, the Sinai billiard, and the hard-sphere gas with periodic boundary conditions or placed in a rectangular box. The periodic orbits that we have in mind are those special orbits where the particles move without collisions or bounce between parallel walls, for example, as illustrated in Fig. 5. These special orbits form sets of zero Lebesgue measure so that they do not prevent the system from being ergodic, or a K-flow, or from having positive average Lyapunov exponents. However, all the Lyapunov exponents of these special periodic orbits vanish because perturbed trajectories may separate from the periodic orbit in an algebraic way: $\|\Phi^t(\mathbf{X}_{\text{p.o.}} + \delta \mathbf{X}) - \Phi^t(\mathbf{X}_{\text{p.o.}})\| \sim t$. As a consequence, if the system is open and the domain $\mathcal{B}$ contains such a marginally unstable periodic orbit, the decay in Eq. (86) is nonexponential as

$$\nu_0[\Upsilon_{\mathcal{B}}^{(+)}(t)] \sim \frac{1}{t^{Nf-1}} \qquad (105)$$

where $Nf$ is the total number of degrees of freedom of the system, and the escape rate vanishes [17,38,39]. (We note that this algebraic decay results from purely geometric effects - such as appear when particles travel down long corridors without hitting anything- and thus has no direct connection to the long-time tails which appear in the time correlation function expressions for transport coefficients [40,41].)

The large-deviation formalism is very useful in this context because the pressure function is still non-trivial in this case. If the Lyapunov exponents vanish in some regions of phase space, some of the dispersion rates $u(\mathbf{Y})$ may be vanishing for a few values of $\mathbf{Y}$ in the $(\varepsilon, T)$-separated subsets. For large and positive values of $\beta$, the partition function (94) is dominated by the few terms with $u(\mathbf{Y}) = 0$ because all the other terms are exponentially vanishing in the limit $T \to \infty$. As a result, the pressure function is equal to zero for large values of $\beta$. On the other hand, for negative values of $\beta$, the terms with nonvanishing dispersion rates $u(\mathbf{Y})$ dominate the sum in the partition function so that the pressure is then positive and nontrivial. There exists a critical, lowest value of $\beta$ above which the pressure vanishes.

$$P(\beta) = 0 \quad \text{for} \quad \beta_{\text{c}} \leq \beta \qquad (106)$$

Schematic pressure functions are depicted in Figs. 2c, 2d for nonhyperbolic closed and open systems. For closed systems, we observe that the pressure is zero above $\beta = 1$ which is in agreement with the vanishing of the escape rate due to Eq. (105), so that the critical value is $\beta_{\text{c}} = 1$. Moreover, the sum of positive Lyapunov exponents must be defined as the left-sided derivative of the pressure at $\beta = 1$ in the case of a closed system.

The discontinuity in the shape of the pressure function is the evidence of a phenomenon of dynamical phase transitions – so called in analogy with statistical mechanics [7,28–30]. We have the following interpretation. The system is described by the continuous family of invariant measures, $\mu_\beta$. The parameter $\beta$ acts like a filtering parameter. When $\beta > \beta_{\text{c}}$, the measure is concentrated on the regular trajectories which have vanishing Lyapunov exponents. For these measures, we can talk about an ordered or regular phase since the corresponding invariant states are ordered. On the other hand, when $\beta < \beta_{\text{c}}$, the measure $\mu_\beta$ gives dominant probability weights to the nonperiodic trajectories which are uncountable. In this case, we can speak of a chaotic or a disordered phase. Tuning the parameter $\beta$ therefore reveals the chaotic features of the dynamics. This is currently done when we refer to the topological pressure per unit time, i.e., $P(\beta = 0)$, as an indicator of chaos.

However, if the pressure function is analytic away from critical points, it has the further advantage that it can be extrapolated from below criticality up to the value at $\beta = 1$ so as to define a supercritical measure, for instance, in the case of open nonhyperbolic systems. In this way, it is possible to define an effective escape rate, $\gamma_{\text{eff}}$, as well as an effective value for the sum of positive Lyapunov exponents at $\beta = 1$. The effective rate can be evidenced in numerical simulations from the transient behavior of the decay function Eq. (86). In many nonhyperbolic systems with a large number of degrees of freedom, the power law decay Eq. (105) may remain a very small effect which is visible only after extremely long times because marginally unstable periodic orbits are very rare and the decay could appear exponential within statistical errors. In nonhyperbolic systems, although we may know from theoretical arguments that the escape rate defined by Eq. (87) actually vanishes, the concept of an effective escape rate is useful to characterize in a rigorous way a numerically observed exponential decay.

## III. THE HARD-SPHERE GAS

In this section we will outline the methods that must be used to apply the considerations of section II to a gas of hard spheres. The treatment of the hard-sphere gas as a dynamical system requires a different analysis than that given in section II for smooth dynamical systems or in section IV for stochastic dynamical systems. Sinai, Bunimovich, Chernov and others [14,15] have developed a very beautiful description of billiard systems in order to analyze their ergodic properties. Here we will use the methods of Sinai *et al.* in order accomplish two goals: (1) To present the Sinai method in a somewhat elementary way in order to acquaint a more physically oriented audience with these techniques which we believe are very useful for the description of the dynamics of hard-sphere systems; and (2) To show how the sum of the positive Lyapunov exponents can be expressed in terms of the average value of the trace of a certain matrix, called the second fundamental operator or curvature matrix. With this result one can begin to apply the escape-rate method to hard-sphere systems.

In this section we will discuss the theory of billiard systems from the point of view of geometrical optics by studying ray trajectories and perturbations thereof. Various notions of elementary differential geometry will be important here, and the reader is advised to consult [42] for an introduction to this subject. We also mention a recent, and extensive review of billiard systems by Tabachnikov [43].

### A. Definition of the billiard

The hard-sphere gas can be thought of as a generalized billiard in a configuration space of dimension $Nf$ where $N$ is the number of hard spheres, each of mass $m$ and diameter $d$, and $f$ is the spatial dimension of each "sphere", i.e. $f = 2$ for disks, $f = 3$ for spheres, etc. The positions and velocities of all of the spheres are given by the $Nf$ dimensional vectors $\mathbf{q}$, and $\mathbf{u}$,

$$\mathbf{q} = (\mathbf{r}_1, \mathbf{r}_2, \mathbf{r}_3, ..., \mathbf{r}_N) \tag{107}$$

and

$$\mathbf{u} = (\mathbf{v}_1, \mathbf{v}_2, \mathbf{v}_3, ..., \mathbf{v}_N) \tag{108}$$

respectively, where $\mathbf{r}_i$, and $\mathbf{v}_i$ are the position and velocity of particle $i$. The Hamiltonian for the system is

$$H = \sum_{i=1}^{N} \frac{1}{2} m \mathbf{v}_i^2 . \tag{109}$$

The hard-sphere condition requires that

$$|\mathbf{r}_i - \mathbf{r}_j| \geq d \qquad \text{for } i \neq j \tag{110}$$

and that the particles undergo instantaneous collisions when $|\mathbf{r}_i - \mathbf{r}_j| = d$. Since energy is conserved in elastic collisions, the magnitude of the velocity $\mathbf{u}$ remains unchanged in a collision. We can then rescale the velocity to a unit value $\mathbf{u}^2 = 1$. Then up to a rescaling factor, the dynamics is the same on every energy shell, $H = E$.

The positions at which the particles collide $|\mathbf{r}_i - \mathbf{r}_j| = d$ defines a hypersurface $\partial \mathcal{Q}$ in the $Nf$ dimensional space which is the border of the billiard $\mathcal{Q}$. We denote by $\mathbf{n}(\mathbf{q})$ a unit vector which is perpendicular to the hypersurface $\partial \mathcal{Q}$ at the point of impact $\mathbf{q}$ and which is directed inside the billiard $\mathcal{Q}$. If $\sigma(\mathbf{q}) = 0$ is the equation of the hypersurface $\partial \mathcal{Q}$, the normal vector

$$\mathbf{n}(\mathbf{q}) = \frac{\partial_{\mathbf{q}} \sigma}{|\partial_{\mathbf{q}} \sigma|} , \tag{111}$$

defines the linear subspace $\Im$ tangent to the collision hypersurface at $\mathbf{q}$.

Fig. 6 schematically depicts the geometry of an elastic collision. If $\mathbf{u}^{(-)}$ is the velocity before the collision, the velocity after the collision is such that the component in the plane $\Im$ tangent to the hypersurface at $\mathbf{q}$ remains unchanged while the normal component changes sign, i.e.,

$$\mathbf{u}^{(+)} = \mathbf{u}^{(-)} - 2(\mathbf{n} \cdot \mathbf{u}^{(-)})\mathbf{n} \tag{112}$$

which is the rule of elastic collisions (and the basic equation of geometrical optics).

For the following discussion we also need to define the linear subspaces $\Im^{(\pm)}$ which are perpendicular to the velocities $\mathbf{u}^{(\pm)}$. Following Sinai, we introduce the transformations:

(a) The projection of $\Im^{(+)}$ on to $\Im^{(-)}$, parallel to $\mathbf{n}$,

$$\mathbf{U}: \quad \Im^{(+)} \to \Im^{(-)}, \qquad \text{given by} \qquad \mathbf{U} = \mathbf{1} - 2\,\mathbf{n}\,\mathbf{n}^{\mathrm{T}}, \tag{113}$$

which takes a vector in the hyperplane $\Im^{(+)}$ and projects it onto the plane $\Im^{(-)}$ such that the difference between the two vectors is parallel to $\mathbf{n}$, and that the components of the two vectors in the directions perpendicular to $\mathbf{n}$ are equal. Here the superscript T denotes a transposed vector, or operator, and $\mathbf{1}$ is a unit operator. The transformation $\mathbf{U}$ is an invertible, orthogonal transformation given by an orthogonal matrix in the $Nf$ dimensional configuration space

$$\mathbf{U} = \mathbf{U}^{\mathrm{T}} = \mathbf{U}^{-1} = \mathbf{1} - 2\,\mathbf{n}\,\mathbf{n}^{\mathrm{T}}, \tag{114}$$

with determinant $\det \mathbf{U} = -1$.

(b) The projection of $\Im^{(-)}$ onto $\Im$ parallel to $\mathbf{u}^{(-)}$

$$\mathbf{V}: \quad \Im^{(-)} \to \Im,$$

is given by the matrix

$$\mathbf{V} = \mathbf{1} - \frac{\mathbf{u}^{(-)}\,\mathbf{n}^{\mathrm{T}}}{\mathbf{u}^{(-)} \cdot \mathbf{n}}. \tag{115}$$

(c) The projection of $\Im$ onto $\Im^{(-)}$, parallel to $\mathbf{n}$

$$\mathbf{V}^{\mathrm{T}}: \Im \to \Im^{(-)} \tag{116}$$

is given by the transpose of the matrix $\mathbf{V}$, Eq. (115).

### B. The Second Fundamental Form

The convexity of the hypersurface $\sigma(\mathbf{q}) = 0$ determines the defocusing character of the collisions. The convexity is characterized by an operator known as the second fundamental form,[4] and which gives the variation of the normal vector with respect to a variation of a point on the hypersurface. That is, we define the second fundamental form $\mathbf{K}(\mathbf{q})$ by

$$\mathbf{n}(\mathbf{q}+\delta\mathbf{q}) - \mathbf{n}(\mathbf{q}) = \mathbf{K}(\mathbf{q}) \cdot \delta\mathbf{q} \tag{117}$$

where both $\mathbf{q}$ and $\mathbf{q}+\delta\mathbf{q}$ belong to the hypersurface. Consequently, the variation $\delta\mathbf{q}$ is perpendicular to the normal, $\mathbf{n} \cdot \delta\mathbf{q} = 0$. Similarly, $\mathbf{n}^2 = 1$ implies that $\mathbf{n} \cdot \delta\mathbf{n} = 0$. As a result, the second fundamental form maps vectors, $\delta\mathbf{q}$, in the tangent plane onto vectors $\delta\mathbf{n}$ in the tangent plane,

$$\mathbf{K}: \Im \to \Im, \tag{118}$$

which is symmetric. If the operator is nonnegative, $\mathbf{K} \geq 0$, the collisions are defocusing or neutral, and they may lead to a dynamical instability. In the case that $\mathbf{K} \leq 0$ the collisions are focusing or neutral, but, as illustrated by

---

[4] The first fundamental form determines the Riemannian distance between two nearby points on the surface.

Bunimovich's stadium billiard, the collisions do not necessarily lead to a dynamically stable situation. In order to see the meaning of the sign of the second fundamental form consider the inner product

$$\delta \mathbf{q}^T \cdot \delta \mathbf{n} = \delta \mathbf{q}^T \cdot \mathbf{K} \cdot \delta \mathbf{q} . \tag{119}$$

For surfaces that are defocusing, this inner product satisfies $\delta \mathbf{q}^T \cdot \delta \mathbf{n} > 0$, while for focusing surfaces $\delta \mathbf{q}^T \cdot \delta \mathbf{n} < 0$ as may be immediately checked by considering a sphere in three dimensions and constructing this inner product for both outward ($\delta \mathbf{q}^T \cdot \delta \mathbf{n} > 0$) and inward ($\delta \mathbf{q}^T \cdot \delta \mathbf{n} < 0$) normals. In the case of outward normals,

$$\mathbf{K} = \frac{1}{d} \mathbf{1}_\perp$$

where $d$ is the radius of the sphere, $\mathbf{1}_\perp = \mathbf{e}_\theta \mathbf{e}_\theta^T + \mathbf{e}_\phi \mathbf{e}_\phi^T$ is the unit operator in the tangent plane to the sphere at point $\mathbf{q}$, and $\mathbf{e}_\theta$ and $\mathbf{e}_\phi$ are mutually orthogonal unit vectors in this tangent plane.

### C. An Example

We now construct the various geometric quantities that describe a collision between particles 1 and 2. The hypersurface of collision is locally defined by

$$|\mathbf{r}_1 - \mathbf{r}_2|^2 = (x_1 - x_2)^2 + (y_1 - y_2)^2 + (z_1 - z_2)^2 = d^2 . \tag{120}$$

A parametric representation of this hypersurface is given in terms of center of mass and relative coordinates by

$$\mathbf{q} = (\mathbf{R}_{12} + \frac{d}{2}\boldsymbol{\epsilon}_{12}, \mathbf{R}_{12} - \frac{d}{2}\boldsymbol{\epsilon}_{12}, \mathbf{r}_3, \mathbf{r}_4, ..., \mathbf{r}_N) \tag{121}$$

where $\mathbf{R} = (\mathbf{r}_1 + \mathbf{r}_2)/2$ is the location of the center of mass of the two spheres 1 and 2 at the instant of the collision and $\boldsymbol{\epsilon}_{12}$ is the unit vector in the direction of the line joining their centers at the time of the collision. The unit vector which is normal to the hypersurface at the position Eq. (121) is given by

$$\mathbf{n} = \frac{1}{\sqrt{2}}(\boldsymbol{\epsilon}_{12}, -\boldsymbol{\epsilon}_{12}, \mathbf{0}, ..., \mathbf{0}). \tag{122}$$

After applying the collision rule, Eq. (112) we find the usual result

$$\begin{aligned}
\mathbf{v}_1^{(+)} &= \mathbf{v}_1^{(-)} - [\boldsymbol{\epsilon}_{12} \cdot (\mathbf{v}_1^{(-)} - \mathbf{v}_2^{(-)})]\boldsymbol{\epsilon}_{12} \\
\mathbf{v}_2^{(+)} &= \mathbf{v}_2^{(-)} + [\boldsymbol{\epsilon}_{12} \cdot (\mathbf{v}_1^{(-)} - \mathbf{v}_2^{(-)})]\boldsymbol{\epsilon}_{12} \\
\mathbf{v}_3^{(+)} &= \mathbf{v}_3^{(-)} \\
&\vdots \\
\mathbf{v}_N^{(+)} &= \mathbf{v}_N^{(-)}
\end{aligned} \tag{123}$$

which conserves the total energy and momentum.

The defocusing character of hard-sphere collisions is shown by considering the second fundamental form. The variation in position on the hypersurface is achieved by making a change in the unit vector $\boldsymbol{\epsilon}_{12}$ and in $\mathbf{r}_3, ..., \mathbf{r}_N$. Thus $\delta \mathbf{q}$ is given by

$$\delta \mathbf{q} = (\frac{d}{2}\delta\boldsymbol{\epsilon}_{12}, -\frac{d}{2}\delta\boldsymbol{\epsilon}_{12}, \delta\mathbf{r}_3, ..., \delta\mathbf{r}_N) , \tag{124}$$

and $\delta \mathbf{n}$ is

$$\delta \mathbf{n} = \frac{1}{\sqrt{2}}(\delta\boldsymbol{\epsilon}_{12}, -\delta\boldsymbol{\epsilon}_{12}, \mathbf{0}, ..., \mathbf{0}) . \tag{125}$$

From these two results it follows immediately that the second fundamental form is given as

$$\mathbf{K} = \frac{\sqrt{2}}{d}(\mathbf{1}, \mathbf{1}, \mathbf{0}, \mathbf{0}, ..., \mathbf{0}) . \tag{126}$$

Clearly the collision between the two spheres is defocusing as expected. However, defocusing only occurs in two directions while the other directions are neutral so that such billiards are referred to as semidispersing billiards [14]. One can think of the $Nf$ dimensional configuration space for hard spheres as being bounded by hypercylinders with both defocusing and neutral directions much like ordinary three dimensional cylinders.

## D. Linear Stability

To develop a method to compute dynamical quantities for a hard-sphere gas we need to consider infinitesimal perturbations of some reference trajectory. The perturbations are best represented in a local frame of coordinates with one axis parallel to the velocity $\mathbf{u}$

$$\delta \mathbf{q} = \delta q_\parallel \mathbf{e}_\parallel + \delta q_{\perp 1} \mathbf{e}_{\perp 1} + \delta q_{\perp 2} \mathbf{e}_{\perp 2} + \cdots \delta q_{\perp Nf-1} \mathbf{e}_{\perp Nf-1} \tag{127}$$

where $\mathbf{e}_\parallel = \mathbf{u}$ with $\mathbf{u} \cdot \mathbf{u} = 1$. An equation similar to Eq. (127) obtains for the infinitesimal variation in velocity with respect to the reference trajectory. We remark that the condition $\mathbf{u}^2 = 1$ implies that the velocity perturbation is necessarily perpendicular to $\mathbf{u}$: $\mathbf{u} \cdot \delta \mathbf{u} = 0$. Therefore we have $\delta u_\parallel = 0$. Similarly we can set $\delta q_\parallel = 0$, otherwise the perturbation can be assigned to another position along the trajectory. Consequently, the infinitesimal perturbation belongs to a linear space of dimension $2(Nf - 1)$.

Our purpose is to obtain the time evolution of the infinitesimal perturbation over a time interval $T$ which will define the monodromy matrix $\mathsf{M}(T)$

$$\begin{pmatrix} \delta \mathbf{q}_\perp(T) \\ \delta \mathbf{u}_\perp(T) \end{pmatrix} = \mathsf{M}(T) \cdot \begin{pmatrix} \delta \mathbf{q}_\perp(0) \\ \delta \mathbf{u}_\perp(0) \end{pmatrix} . \tag{128}$$

The time evolution is composed of collisions and free flights between the collisions. We shall first determine the monodromy matrix for a free flight and then the matrix for a collision.

## E. Free Flight

During a free flight the velocity vector is constant although the position changes according to

$$\mathbf{q}(t) = \mathbf{u}(0) \ t + \mathbf{q}(0) , \tag{129}$$

where $[\mathbf{q}(0), \mathbf{u}(0)]$ are the position and velocity after the previous collision. As a direct consequence, before the next collision

$$\begin{pmatrix} \delta \mathbf{q}_\perp(t) \\ \delta \mathbf{u}_\perp(t) \end{pmatrix} = \begin{pmatrix} \delta \mathbf{q}_\perp(0) + t \ \delta \mathbf{u}_\perp(0) \\ \delta \mathbf{u}_\perp(0) \end{pmatrix} \tag{130}$$

where $t$ is the time of flight between two consecutive collisions, which is also equal to the distance between impact points since the velocity $\mathbf{u}$ is normalized to unity. The monodromy matrix for a free flight is therefore

$$\mathsf{M}_{\text{free flight}} = \begin{pmatrix} \mathbf{1} & t \, \mathbf{1} \\ \mathbf{0} & \mathbf{1} \end{pmatrix} . \tag{131}$$

## F. Collision

If the collision illustrated in Fig. 6 is perturbed, the position and velocity just before the collision are displaced by $(\delta \mathbf{q}_\perp^{(-)}, \delta \mathbf{u}_\perp^{(-)})$ which both belong to the subspace $\Im^{(-)}$ perpendicular to the incident velocity $\mathbf{u}^{(-)}$. As a consequence, the collision of the perturbed trajectory does not occur at the collision point $\mathbf{q}$ of the reference trajectory but at the nearby point $\mathbf{q} + \delta \mathbf{q}$ on the hypersurface. we note that the perturbation $\delta \mathbf{q}$ of the impact point belongs to the tangent subspace $\Im$ and is determined from the perturbed position $\delta \mathbf{q}_\perp^{(-)}$ by the projection from $\Im^{(-)}$ onto $\Im$ parallel to the incident velocity $\mathbf{u}^{(-)}$, that is by Eq. ( 115)

$$\delta \mathbf{q} = \mathsf{V} \cdot \delta \mathbf{q}_\perp^{(-)}. \tag{132}$$

The perturbed trajectory after the collision is issued from the point $\mathbf{q} + \delta \mathbf{q}$ with a velocity $\mathbf{u}^{(+)} + \delta \mathbf{u}^{(+)}$ given by the collison rule Eq. (112). The intersection of this trajectory with the tangent space $\Im$ defines the perturbation in position $\delta \mathbf{q}_\perp^{(+)}$ of the outgoing trajectory. Therefore the outgoing perturbation is given by the projection of the perturbation Eq. (132) of the impact point from the tangent subspace $\Im$ onto the subspace $\Im^{(+)}$ parallel to the outgoing velocity $\mathbf{u}^{(+)}$. In Fig. 6 we observe that the composition of the two successive projections, $\Im^{(-)} \to \Im \to \Im^{(+)}$ is the projection from $\Im^{(-)}$ onto $\Im^{(+)}$ parallel to the normal vector $\mathbf{n}$ which is the inverse $\mathbf{U}^{-1}$ of the isometry Eqs. (113)-(114),

$$\delta \mathbf{q}_\perp^{(+)} = \mathbf{U}^{-1} \cdot \delta \mathbf{q}_\perp^{(-)}. \tag{133}$$

At the perturbed impact point, the normal vector is no longer identical to $\mathbf{n}$ but is perturbed as determined by the second fundamental form, Eq. (118). Accordingly, the velocity perturbation after the collision can be obtained in a straightforward way by differentiating the collision rule Eq. (112)

$$\delta \mathbf{u}_\perp^{(+)} = \delta \mathbf{u}_\perp^{(-)} - 2(\mathbf{n} \cdot \delta \mathbf{u}_\perp^{(-)})\mathbf{n} - 2(\mathbf{n} \cdot \mathbf{u}^{(-)})\mathbf{K} \cdot \delta \mathbf{q} - 2(\mathbf{u}^{(-)} \cdot \mathbf{K} \cdot \delta \mathbf{q})\mathbf{n} \tag{134}$$

where $\delta \mathbf{q}$ is the perturbation of the impact point, Eq. (132). Defining the angle $\phi$ by

$$\cos \phi = \mathbf{n} \cdot \mathbf{u}^{(+)} = -\mathbf{n} \cdot \mathbf{u}^{(-)} \tag{135}$$

and using Eqs. (114) and (132), we find that Eq. (134) becomes

$$\delta \mathbf{u}_\perp^{(+)} = \mathbf{U}^{-1} \cdot \left[ \delta \mathbf{u}_\perp^{(-)} + (2 \cos \phi) \mathbf{V}^\mathrm{T} \cdot \mathbf{K} \cdot \mathbf{V} \cdot \delta \mathbf{q}_\perp^{(-)} \right]. \tag{136}$$

According to Eqs. (133) and (136), the monodromy matrix of a collision is

$$\mathbf{M}_\mathrm{collision} = \begin{pmatrix} \mathbf{U}^{-1} & \mathbf{0} \\ (2\cos\phi)\mathbf{U}^{-1} \cdot \mathbf{V}^\mathrm{T} \cdot \mathbf{K} \cdot \mathbf{V} & \mathbf{U}^{-1} \end{pmatrix} \tag{137}$$

$$= \begin{pmatrix} \mathbf{U}^{-1} & \mathbf{0} \\ \mathbf{0} & \mathbf{U}^{-1} \end{pmatrix} \begin{pmatrix} 1 & \mathbf{0} \\ (2\cos\phi)\mathbf{V}^\mathrm{T} \cdot \mathbf{K} \cdot \mathbf{V} & 1 \end{pmatrix}.$$

### G. Expanding and Contracting Horospheres

The local Lyapunov exponents are determined by the rate of dispersion of trajectories in the vicinity of the reference trajectory. The dispersion is given by the "horosphere" which is a local sphere tangent to a front of trajectories accompanying the reference trajectory and issued from a common initial position in the past. The front is expanding so that we talk about the expanding horosphere which is nothing else than the local unstable manifold. For the case of one particle moving is a fixed array of scatterers this expanding horosphere is easy to describe, and it has been discussed in detail in Refs. [14,17].

The expanding front has a local curvature which is characterized by the second fundamental operator $\mathbf{B}_u$ defined by

$$\delta \mathbf{u}_\perp = \frac{d}{dt} \delta \mathbf{q}_\perp = \mathbf{B}_u \cdot \delta \mathbf{q}_\perp. \tag{138}$$

Accordingly, the local unstable manifold has the following parametric representation

$$\begin{pmatrix} \delta \mathbf{q}_\perp \\ \delta \mathbf{u}_\perp \end{pmatrix} = \begin{pmatrix} \delta \mathbf{q}_\perp \\ \mathbf{B}_u \cdot \delta \mathbf{q}_\perp \end{pmatrix} \tag{139}$$

in the linear subspace. A similar representation holds for the local stable manifold.

Let us consider a trajectory from an initial condition $\mathbf{X} = (\mathbf{q}, \mathbf{u})$. Fig. 7 depicts the backward and forward portions of it. Collisions occur at the impact points $n$ and times $t_n$ with $n \in \mathbb{Z}$. We denote by $\tau_{n+1,n}$ the time between the collisions $n$ and $n+1$. Furthermore, $\mathbf{B}_u^{(-)}(n)$ denotes the second fundamental operator of the horosphere immediately before the $n^\mathrm{th}$ collision, i. e., at the end of the preceding free flight between the collisions $n$ and $n-1$. On the other hand, $\mathbf{B}_u^{(+)}(n)$ denotes the fundamental operator immediately after the $n^\mathrm{th}$ collision. $\mathbf{B}_u(t)$ denotes the fundamental operator during a free flight.

We assume that the second fundamental operator of the expanding horosphere is fixed at some collision in the remote past and look for the operator at the initial condition. The operator is successively modified by the free flight and collisions according to the monodromy matrices Eqs. (131) and (137). The second fundamental operator $\mathbf{B}_u'$ is related to the one before the monodromy matrix $\mathbf{M}$ by

$$\begin{pmatrix} \delta \mathbf{q}'_\perp \\ \delta \mathbf{u}'_\perp \end{pmatrix} = \begin{pmatrix} \delta \mathbf{q}'_\perp \\ \mathbf{B}'_u \cdot \delta \mathbf{q}'_\perp \end{pmatrix} = \mathbf{M} \cdot \begin{pmatrix} \delta \mathbf{q}_\perp \\ \mathbf{B}_u \cdot \delta \mathbf{q}_\perp \end{pmatrix} \tag{140}$$

which is solved by eliminating $\delta \mathbf{q}_\perp$ and $\delta \mathbf{q}'_\perp$ between both lines.

Applying this equation to the monodromy matrix Eq. (131) for a free flight, we find

$$\mathbf{B}_u^{(-)}(n+1) = \left[\tau_{n+1,n}\mathbf{1} + \mathbf{B}_u^{(+)}(n)^{-1}\right]^{-1}, \qquad \text{(free flight)}. \tag{141}$$

For a collision, we have from Eq. (137)

$$\mathbf{B}_u^{(+)}(n) = \mathbf{U}_n^{-1} \cdot \left[(2\cos\phi_n)\mathbf{V}_n^{\mathrm{T}} \cdot \mathbf{K}_n \cdot \mathbf{V}_n + \mathbf{B}_u^{(-)}(n)\right] \cdot \mathbf{U}_n \qquad \text{(collision)}. \tag{142}$$

Combining Eqs. (141) and (142) for successive backward collisions, we obtain Sinai's matrix continued fraction expression for $\mathbf{B}_u(t)$:

$$\mathbf{B}_u(t) =$$

$$= \cfrac{1}{\tau\mathbf{1} + \mathbf{U}_n^{-1} \cdot \left[(2\cos\phi_n)\mathbf{V}_n^{\mathrm{T}} \cdot \mathbf{K}_n \cdot \mathbf{V}_n + \cfrac{1}{\tau_{n,n-1}\mathbf{1} + \mathbf{U}_{n-1}^{-1} \cdot \left[(2\cos\phi_{n-1})\mathbf{V}_{n-1}^{\mathrm{T}} \cdot \mathbf{K}_{n-1} \cdot \mathbf{V}_{n-1} + \cfrac{1}{\tau_{n-1,n-2}\mathbf{1} + \cdots}\right] \cdot \mathbf{U}_{n-1}}\right] \cdot \mathbf{U}_n} \tag{143}$$

with $\tau = t - t_n$ and for $t_{n+1} > t > t_n$. Because $\mathbf{K}_n \geq 0$, and $\cos\phi_n \geq 0$, the matrix $\mathbf{B}_u(t) \geq 0$ so that the expanding character is maintained during the whole time evolution in the case of a hard-sphere gas. Let us emphasize that the operator (143) is defined locally for each initial condition $\mathbf{X}$ and can be obtained by integrating backward the trajectory $\Phi^t \mathbf{X}$ for $-\infty < t < 0$ to determine the successive past collisions and the corresponding quantities appearing in Eq. (143).

To see the connection between the second fundamental operator and the Lyapunov exponents of the hard-sphere gas we proceed as follows. We use the fact that between collisions the quantity $\delta \mathbf{q}_\perp$ develops as

$$\delta \mathbf{q}_\perp(t_n) = \delta \mathbf{q}_\perp(t_{n-1} + \tau_{n,n-1}) = \delta \mathbf{q}_\perp(t_{n-1}) + \tau_{n,n-1}\delta \mathbf{u}_\perp(t_{n-1}) = \left[\mathbf{1} + \tau_{n,n-1}\mathbf{B}_u^{(+)}(n-1)\right] \cdot \delta \mathbf{q}_\perp(t_{n-1}) \tag{144}$$

where the quantities are the values immediately after the collisions, as given by Eq. (142). After a sequence of $n$ collisions, $\delta \mathbf{q}_\perp(t)$ is given by

$$\delta \mathbf{q}_\perp(t_n) = \left[\mathbf{1} + \tau_{n,n-1}\mathbf{B}_u^{(+)}(n-1)\right] \cdot \left[\mathbf{1} + \tau_{n-1,n-2}\mathbf{B}_u^{(+)}(n-2)\right] \cdots \left[\mathbf{1} + \tau_{1,0}\mathbf{B}_u^{(+)}(0)\right] \cdot \delta \mathbf{q}_\perp(0), \tag{145}$$

with $t_0 = 0$. If there is an exponential separation of trajectories we would expect that for long times $||\delta \mathbf{q}_\perp(T)|| \approx (\exp \lambda T)||\delta \mathbf{q}(0)||$. The exponential growth factors $\lambda$ would satisfy the relation

$$\sum_{\lambda_i > 0} \lambda_i = \lim_{T \to \infty} \frac{1}{T} \ln \det\left[\mathbf{1} + \tau_{n,n-1}\mathbf{B}_u^{(+)}(n-1)\right] \cdots \left[\mathbf{1} + \tau_{10}\mathbf{B}_u^{(+)}(0)\right] \tag{146}$$

$$= \lim_{T \to \infty} \frac{1}{T} \sum_{i=0}^{n-1} \ln \det\left[\mathbf{1} + \tau_{i+1,i}\mathbf{B}_u^{(+)}(i)\right]$$

$$= \lim_{T \to \infty} \frac{1}{T} \sum_{i=0}^{n-1} \int_0^{\tau_{i+1,i}} d\tau \, \mathrm{tr}\left[\tau\mathbf{1} + \mathbf{B}_u^{(+)}(i)^{-1}\right]^{-1} \tag{147}$$

Now by using Eq. (141) in the form

$$\mathbf{B}_u(t_n + \tau) = \mathbf{B}_u^{(+)}(n)\left[\mathbf{1} + \tau\mathbf{B}_u^{(+)}(n)\right]^{-1}$$

we readily find that

$$\sum_{\lambda_i > 0} \lambda_i = \lim_{T \to \infty} \frac{1}{T} \int_0^T dt \, \mathrm{tr}\, \mathbf{B}_u(t). \tag{148}$$

In the event that the system is closed and ergodic, then the time average in Eq. (148) can be replaced by an ensemble average with respect to the equilibrium invariant measure of the system, so that in this case

$$\sum_{\lambda_i > 0} \lambda_i = \mu_{\text{eq}}\left(\text{tr}\,\mathbf{B}_u\right) \tag{149}$$

for almost all trajectories. For open systems, a similar result obtains but the appropriate nonequilibrium measure must be used in computing the average of the trace of the second fundamental operator. We can therefore identify the local dispersion rate for billiards as

$$u(\mathbf{X}) = \sum_{\lambda_i > 0} \chi_i(\mathbf{X}) = \text{tr}\,\mathbf{B}_u(\mathbf{X}) \tag{150}$$

The form Eq. (150) plays for hard spheres the role of Eq. (32) for systems with smooth potentials. With respect to smooth Hamiltonian systems, a simplification for billiard systems comes from the fact that the local Lyapunov exponents are given directly in terms of quantities which can be constructed from successive collisions. It is possible that numerical calculations of the sum of positive Lyapunov exponents for billiard systems could be calculated efficiently and quickly using Eqs. (146) or (148). It is worth noting that the individual Lyapunov exponents may also be calculated by using the second fundamental operator (143).

For the special case of a billiard system consisting of one moving particle in a regular triangular array of fixed hard disk scatterers – the triangular Lorentz gas – one can evaluate the continued fraction, Eq. (143), in terms of scalar quantities, and the positive Lyapunov exponent has been determined numerically for both open and closed systems by Gaspard and Baras [16,17]. For open systems, the positive Lyapunov exponent describes the expanding manifold for the trajectories on the fractal repeller. In the case of random Lorentz gases – where a particle moves in a fixed but random array of hard disk or hard-sphere scatterers – van Beijeren, Dorfman, and Latz [18,44] have determined the sum of the positive Lyapunov exponents for both closed and open systems in the case that the density of scatterers is low.

As a consequence of the considerations presented here we see that even for billiard systems it is possible to define the sum of the positive Lyapunov exponents in terms of the average value of the trace of the second fundamental operator. For closed systems, this quantity determines the Kolmogorov-Sinai entropy of the billiard, by Pesin's theorem. For open systems, this quantity provides the sum of the positive Lyapunov exponents for trajectories on the repeller, which is an essential ingredient needed for the escape-rate formalism for open systems.

### H. Marginally Unstable Periodic Orbits

As we remarked in subsection II.F and as is illustrated in Fig. 5, the hard-sphere gas with periodic boundary conditions or in a rectangular box presents special periodic orbits for which the particles do not undergo collisions. As a result the second fundamental operator Eq. (143) is of free-flight type, Eq. (141) which decays to zero as $t^{-1}$ as $t \to \infty$ because $\mathbf{K} = 0$ for the particular case of collisions on flat walls. Therefore, all of the Lyapunov exponents of those special orbits vanish and the system is nonhyperbolic. These special periodic orbits form continuous families of dimension $Nf - 1$. We discussed in subsection II.E how the large-deviation formalism can be applied to such systems.

Since the transport coefficients are bulk properties obtained after a thermodynamic limit, the theory of paper I is also valid for a system which is modified at its boundaries. Using the independence of the transport coefficients of surface effects, we can consider a hard-sphere gas in a box with convex walls as in Fig. 8. Fig. 8a shows walls given by portions of spheres with a radius of order of the box size. In this case, the convexity is global and mild but enough to turn the system into a hyperbolic system. Indeed, the families of special orbits have now disappeared and all of the periodic orbits are now unstable.

Another possibility is shown in Fig. 8b where the walls are composed of many portions of small spheres modeling the atoms of the walls. Here also, the continuous families of special periodic orbits have disappeared. There may still remain rare periodic orbits where the particles have no mutual collisions but these periodic orbits are now unstable since the walls are defocusing. Therefore, the system here is also hyperbolic and the escape rate formula Eq. (100) strictly applies.

## IV. LATTICE-GAS AUTOMATA

### A. Lattice-gas automata as Markov chains

Lattice-gas automata are probabilistic Markov chains over system states defined by the set of discrete states taken by individual particles. The single particle state for particle $i$, $\xi_i$, takes one among $m$ integers which determine its position and velocity. The collection of all the single-particle states $(\xi_1, \xi_2, \xi_3, \ldots, \xi_N) = \omega$ defines a state of the whole system. As a consequence the total number of possible system states is $m^N$ [45]. This set of possible system states plays the role of an alphabet $\mathcal{A} = \{1, 2, 3, \ldots, m^N\}$ of a symbolic dynamics $\mathcal{M} = \Sigma_{\mathcal{A}}$ in which the trajectories of the system correspond to bi-infinite sequences of states

$$\boldsymbol{\omega} = \ldots \omega_{-2} \omega_{-1} \cdot \omega_0 \omega_1 \omega_2 \ldots . \tag{151}$$

The successive symbols $\omega_k$ give the state of the system at successive times $t_k = k\Delta t$. The set of all bi-infinite sequences defines the phase space of the lattice-gas automaton. This so-defined phase space is a continuum, as it should be. For example a simple automaton with an alphabet of two symbols, 0 and 1, and a set of trajectories consisting of all bi-infinite sequences of these two sysmbols can be mapped onto a unit square, as is done in the baker's transformation. We remark that a symbolic dynamics establishes a one-to-one correspondence between the system trajectories and the bi-infinite sequences.

Gaspard and Wang have shown elsewhere [46] that an area-preserving map can be constructed which is isomorphic to the Markov chain. In this way we can establish a connection between some properties of the Markov chain and those of a deterministic dynamical system. However, the dimensionality is lost in this connection. As a consequence, geometric properties of dimension which are typical of differentiable dynamical systems cannot be recovered. Nevertheless, we are here interested in the quantities appearing in the escape-rate formalism. In particular, a formula like Eq. (2) only contains the sum of positive Lyapunov exponents, a global quantity which does not require the knowledge of individual Lyapunov exponents. Therefore the escape-rate formalism can be extended to lattice-gas automata.

The lattice-gas automaton is fully characterized by the matrix $\mathbf{P}$ of the Markov chain on the system states. The elements $P_{\omega\omega'}$ of the matrix $\mathbf{P}$ give the probabilities of transition between two successive states $\omega$ and $\omega'$. Since probability is conserved and the Markov chain admits an invariant vector $\{p_\omega\}$ we know that

$$\sum_{\omega'} P_{\omega\omega'} = 1, \qquad \text{and} \qquad \sum_\omega p_\omega P_{\omega\omega'} = p_{\omega'} \tag{152}$$

with $\sum_\omega p_\omega = 1$. We may assume that the Markov chain is ergodic so that the invariant vector is unique. The invariant probability measure is then defined as

$$\mu_{\mathcal{A}}(\omega_0 \omega_1 \ldots \omega_{n-1}) = p_{\omega_0} P_{\omega_0 \omega_1} P_{\omega_1 \omega_2} \ldots P_{\omega_{n-2} \omega_{n-1}}. \tag{153}$$

### B. Escape-rate formalism of Markov chains

Although the dynamics takes place within the full set of states, $\mathcal{A}$, there may exist a subset of states, $\mathcal{B} \subset \mathcal{A}$, which are visited in a transient way. This might happen, for instance, if we impose a nonequilibrium constraint on the system, such as requiring that a Helfand moment lie in an interval Eq. (1). If the initial state belongs to this subset, the state will escape from this subset after a finite time for almost all of the trajectories Eq. (151). Nevertheless there exist trajectories Eq. (151) which remain forever on this subset $\mathcal{B}$. These trapped trajectories are exceptional in the sense that their measure, Eq. (153), is vanishing and consequently they form a repeller $\mathcal{R}_{\mathcal{B}} = \Sigma_{\mathcal{B}}$. The first time at which a given trajectory escapes from the subset $\mathcal{B}$ defines a problem of first passage. The dynamics of the full system with respect to the repeller is similar to a scattering process where the trajectory visits for a while a vicinity of the repeller. In this image where a trajectory is going in and out of the subset $\mathcal{B}$, the repeller may be considered as a predefined nonequilibrium fluctuation, as discussed elsewhere [47,48].

Since the dynamics is transient on the repeller there exists an escape rate $\gamma$. We can calculate this escape rate by constructing the Markov subchain between states of the subset $\mathcal{B}$. The values of the transition probabilities are given by a submatrix which is contained in the matrix $\mathbf{P}$ of the full Markov chain Eq. (152). This submatrix $\mathbf{Q}$ is composed of the elements of $\mathbf{P}$ between the states of the subset $\mathcal{B}$

$$\mathbf{P} = \begin{array}{c} \\ \mathcal{A} \setminus \mathcal{B} \\ \mathcal{B} \end{array} \begin{array}{c} \mathcal{A} \setminus \mathcal{B} \quad \mathcal{B} \\ \begin{pmatrix} * & * \\ * & \mathbf{Q} \end{pmatrix} \end{array}, \tag{154}$$

where $*$ denotes other submatrices.

Because it is obtained by truncating the full matrix $\mathbf{P}$, the submatrix $\mathbf{Q}$ is not a stochastic matrix obeying Eq. (152). In particular, the leading eigenvalue of $\mathbf{Q}$ is no longer 1 but is smaller than 1. Actually the eigenvalues of the submatrix $\mathbf{Q}$ give the decay rates out of the subset $\mathcal{B}$ so that the leading eigenvalue gives the escape rate $\gamma$ according to

$$\mathbf{Q}|v\rangle = \exp(-\gamma)|v\rangle, \qquad \text{and} \qquad \langle u|\mathbf{Q} = \exp(-\gamma)\langle u| \tag{155}$$

which defines the leading right and left eigenvectors of $\mathbf{Q}$. The escape rate is here in units of the time steps, $\Delta t$, of the automaton. If $\mathbf{Q}$ is a nonnegative, irreducible, aperiodic matrix then according to the Perron-Frobenius theorem all of the other eigenvalues are less in absolute value the the largest eigenvalue, $\exp(-\gamma)$.

Using the right and left eigenvectors we can construct the Markov subchain of the repeller dynamics in terms of the following matrix and its associated invariant vector

$$\Pi_{\omega\omega'} = \exp(\gamma) Q_{\omega\omega'} \frac{v_{\omega'}}{v_\omega}, \qquad \text{and} \qquad \pi_\omega = \frac{u_\omega v_\omega}{\langle u|\ v\rangle} \tag{156}$$

which is now stochastic since

$$\sum_{\omega'\in\mathcal{B}} \Pi_{\omega\omega'} = 1, \qquad \sum_{\omega\in\mathcal{B}} \pi_\omega \Pi_{\omega\omega'} = \pi_{\omega'}, \qquad \text{and} \qquad \sum_{\omega\in\mathcal{B}} \pi_\omega = 1. \tag{157}$$

This stochastic matrix defines an invariant measure on the repeller which is given by

$$\mu_\mathcal{B}(\omega_0\omega_1\ldots\omega_{n-1}) = \pi_{\omega_0} \Pi_{\omega_0\omega_1} \Pi_{\omega_1\omega_2} \ldots \Pi_{\omega_{n-2}\omega_{n-1}} \tag{158}$$

where $\omega_k \in \mathcal{B}$. This new invariant measure $\mu_\mathcal{B}$ gives nonvanishing probabilities only for trajectories staying on the repeller.

### C. Lattice-gas automata as chaotic dynamical systems

We now have the necessary elements to proceed with a derivation of the escape-rate formula. The escape rate $\gamma$ was already defined in terms of the leading eigenvalue of the submatrix $\mathbf{Q}$. On the other hand, the KS entropy of the dynamics on the repeller is defined as the KS entropy per time step $\Delta t$ of the Markov chain Eqs. (157) and (158),

$$h_{\text{KS}}(\mu_\mathcal{B}) = -\sum_{\omega,\omega'\in\mathcal{B}} \pi_\omega \Pi_{\omega\omega'} \ln \Pi_{\omega\omega'}. \tag{159}$$

We need to find the quantity which plays the role of the sum of the positive Lyapunov exponents, $\sum_{\lambda_i>0} \lambda_i$. It is not obvious at first sight how to do this. The solution can be found by mapping a Markov chain onto a deterministic map [46,49], and then computing the sum of the positive Lyapunov exponents for the deterministic map. One finds in this way that the inverses of the probabilities, $Q_{\omega\omega'}$, for the separate steps of the Markov chain play the role of the stretching factors by which trajectories separate in the map or by which the probability is dispersed in the Markov chain. The inverse of the dispersion factor for the trajectories visiting successively the states $\omega_0\omega_1\omega_2\ldots\omega_{n-1}$ is therefore given by

$$\exp -U(\omega_0\omega_1\ldots\omega_{n-1}) = Q_{\omega_0\omega_1} Q_{\omega_1\omega_2} \ldots Q_{\omega_{n-2}\omega_{n-1}}. \tag{160}$$

The role of the sum of the Lyapunov exponents is then played by the average quantity

$$\mu_\mathcal{B}(u) = \lim_{n\to\infty} \frac{1}{n} \sum_{\omega_0\ldots\omega_{n-1}\in\mathcal{B}} \mu_\mathcal{B}(\omega_0\ldots\omega_{n-1}) \ln \exp U(\omega_0\ldots\omega_{n-1}), \tag{161}$$

with the definition Eq. (160). Eqs. (160)-(161) define for lattice-gas automata the analogue of the dispersion rates given by Eq. (32) for systems with smooth potentials and by Eq. (150) for hard spheres.

Using the factorization property of the dispersion factors, Eq. (160), and the Markov property of the measure Eqs. (157) and (158), we obtain

$$\mu_\mathcal{B}(u) = \sum_{\omega,\omega'\in\mathcal{B}} \pi_\omega \Pi_{\omega\omega'} \ln \frac{1}{Q_{\omega\omega'}} \tag{162}$$

which is positive since the matrix elements are smaller than one: $Q_{\omega\omega'} < 1$. The dispersion rate $u$ has the same units, $\Delta t^{-1}$, as the escape rate $\gamma$ and the KS entropy, Eq. (159). The isomorphism between the Markov chain and a deterministic, area-preserving map has been discussed in detail for the multibaker transformation and for Lorentz lattice gases [48,49]. This isomorphism shows that the sum of the positive Lyapunov exponents for the deterministic systems is identical to $\mu_\mathcal{B}(u)$ given above,

$$\mu_\mathcal{B}(u) = \sum_{\lambda_i > 0} \mu_\mathcal{B}(\lambda_i) \tag{163}$$

both calculated on the repeller $\Sigma_\mathcal{B}$. This last property guarantees that the averaged quantity $u$ is the unique analog of the sum of the positive mean Lyapunov exponents for lattice-gas automata.

Replacing the definitions Eq. (156) of the matrix $\mathbf{\Pi}$ and of the invariant vector $\pi$ in the expression Eq. (159) for the KS entropy, we obtain the identity

$$\gamma = \mu_\mathcal{B}(u) - h_{\mathrm{KS}}(\mu_\mathcal{B}) \tag{164}$$

which is the escape-rate formula (100) for lattice-gas automata. This formula gives the escape rate from the repeller $\Sigma_\mathcal{B}$ as the difference between an average dispersion rate playing the role of the sum of the positive Lyapunov exponents and the KS entropy.

Similarly, we have the analog of the Pesin theorem. When we relax the constraint on the subset of allowed states, then $\mathcal{A} = \mathcal{B}$, the repeller becomes the full phase space, and $\mathbf{Q}=\mathbf{P}$. As a consequence the escape rate vanishes and

$$\mu_\mathcal{A}(u) = h_{\mathrm{KS}}(\mu_\mathcal{A}). \tag{165}$$

We can also define the dynamical pressure function of section II for lattice-gas automata according to

$$P(\beta) = \lim_{n \to \infty} \frac{1}{n} \ln \sum_{\omega_0, \omega_1, \ldots, \omega_{n-1} \in \mathcal{B}} (Q_{\omega_0 \omega_1} Q_{\omega_1 \omega_2} \ldots Q_{\omega_{n-2} \omega_{n-1}})^\beta. \tag{166}$$

This pressure function has all of the properties described in section II, which relate it to the preceding quantities like the escape rate, the dispersion rate $u$ and the KS entropy

$$\gamma = -P(1), \qquad \mu_\mathcal{B}(u) = -\left.\frac{dP}{d\beta}\right|_{\beta=1}, \qquad \text{and} \qquad h_{\mathrm{KS}}(\mu_\mathcal{B}) = P(1) - \left.\frac{dP}{d\beta}\right|_{\beta=1}. \tag{167}$$

It is a straightforward calculation to show that the quantities $\gamma$, $\mu_\mathcal{B}(u)$, and $h_{\mathrm{KS}}(\mu_\mathcal{B})$ derived from the pressure function are identical with those given by Eqs. (155), (162), and (159), respectively. This is accomplished by using the spectral decomposition of the matrix $\mathbf{Q}$, in terms of its eigenvalues and eigenvectors. If this matrix is irreducible and aperiodic the dominant behavior of $\mathbf{Q}^n$, for large $n$ is determined the the largest eigenvalue and the associated left and right eigenvectors. The correspondence of the two kinds of expressions for the dynamical quantities then follows immediately.

Moreover, the analog of the topological entropy per time step $\Delta t$ is given by

$$h_{\mathrm{top}}(\mathcal{B}) = P(0). \tag{168}$$

The topological entropy is the largest eigenvalue of the topological transition matrix associated with the Markov subchain $\mathbf{\Pi}$, i.e., the matrix with elements 0 or 1 depending upon whether $\Pi_{\omega\omega'}$ is zero or positive. If the subchain matrix $\mathbf{Q}$ is a strictly positive $M_\mathcal{B} \times M_\mathcal{B}$ matrix with $M_\mathcal{B} \leq m^N$, then the topological entropy is equal to $h_{\mathrm{top}}(\mathcal{B}) = \ln M_\mathcal{B} \leq N \ln m$.

## V. RELATIONSHIPS TO OTHER THEORIES AND APPROACHES

Recently, there has been substantial progress in understanding the relationship and connections between a number of approaches for describing transport processes in classical systems on the basis of dynamical systems theory. In this section, we briefly describe this progress and indicate some directions for further work.

## A. Spectral theory of the Perron-Frobenius operator

In classical statistical mechanics, the time evolution of the probability density $\rho$ is described by the Liouville equation

$$\partial_t \rho = \hat{L} \rho \qquad \text{where} \qquad \hat{L} = \{H, \cdot\} \qquad (169)$$

is the Poisson bracket with the Hamiltonian. The solution of Eq. (169) is formally given by $\rho_t = \exp(\hat{L}t)\rho_0$. Alternatively, the time evolution on an energy shell $\mathcal{M}$ is described by the Perron-Frobenius operator

$$\rho_t(\mathbf{X}) = \rho_0(\Phi^{-t}\mathbf{X}) = \int_{\mathcal{M}} \delta(\mathbf{X} - \Phi^t \mathbf{Y}) \rho_0(\mathbf{Y}) \, d\mathbf{Y} \equiv (\hat{P}^t \rho_0)(\mathbf{X}) \qquad (170)$$

This operator has a kernel which is a Schwartz distribution rather than a function, indicative, of course, of the unique time development of a given phase point, $\mathbf{Y}$. This is to be contrasted with the type of kernels appearing in the expressions for the time development of probability densities that characterize stochastic processes, such as in solutions to the Fokker-Planck equation, which involve a standard Green's function. Resolving this difference, that is, showing how deterministic systems can show stochastic-like behavior, is at the heart of the major difficult problems of nonequilibrium statistical mechanics. Recently, some of these difficulties seem to have been overcome with the development of a spectral theory of Perron-Frobenius operators for hyperbolic systems, in particular, by Pollicott and Ruelle [50]. These authors have shown that the Perron-Frobenius operator admits resonances corresponding to exponentially decaying eigenstates

$$\hat{P}^t |\phi_n\rangle = e^{s_n t} |\phi_n\rangle \qquad \text{for} \qquad t > 0 \qquad (171)$$

The eigenstates $\{|\phi_n\rangle\}$ are distributions without any corresponding density which explains the failure of previous attempts to define decaying eigenmodes in terms of density functions. The resonances characterize the forward semigroup so that $\text{Re} s_n < 0$. By time reversibility, antiresonances are associated to each one of them, which characterize the backward semigroup. The forward time evolution of an observable $A$ can be decomposed like

$$\langle A | \hat{P}^t | \rho_0 \rangle = \sum_n \langle A | \phi_n \rangle \, e^{s_n t} \, \langle \tilde{\phi}_n | \rho_0 \rangle + \ldots \qquad (172)$$

where the dots denote other terms due to a possible Jordan-block structure [51–53]. The eigenstates $\{|\phi_n\rangle\}$ and their adjoint $\{\langle \tilde{\phi}_n|\}$ are distributions which are defined on sufficiently smooth observables $A$ and initial densities $\rho_0$.

For closed systems which are ergodic and mixing, the leading resonance is vanishing $s_0 = 0$ and it corresponds to the unique microcanonical invariant measure so that $\langle A | \phi_0 \rangle = \mu_{\text{eq}}(A)$ and $\langle \tilde{\phi}_0 | \rho_0 \rangle = \mu_{\text{eq}}(\rho_0)$. The higher resonances give the relaxation rates toward equilibrium in the forward semigroup.

For scattering systems which are decaying, the leading resonance gives the escape rate $s_0 = -\gamma$ [54,53]. The corresponding eigenstate $|\phi_0\rangle$ is concentrated on the unstable manifolds of the repeller: $\text{Cl}[W_u(\mathcal{R})]$. Such a state corresponds to the construction of the fractal by the process Eq. (80) and the definition of the escape rate by Eq. (88). By time reversibility, the values of the escape rate obtained in both theories are identical according to Eqs. (87)-(89).

In the context of transport phenomena, one considers a spatially extended system, which often can be suitably described by imposing appropriate periodic boundary conditions [53,55]. The dynamics over the full phase space can then be reduced to the dynamics inside a basic cell of the system by spatial Fourier transforms which introduce the wavenumber $\mathbf{k}$. The density can be decomposed into components $\rho_{\mathbf{k}}$ of wavenumber $\mathbf{k}$ which are quasiperiodic in space according to Bloch's theorem. These components evolve in time under the following Perron-Frobenius operator which now depends on the wavenumber $\mathbf{k}$ [56]

$$(\hat{P}^t_{\mathbf{k}} \rho_{\mathbf{k}})(\mathbf{X}) = \int e^{i\mathbf{k}\cdot\mathbf{a}(\mathbf{X})} \delta(\mathbf{X} - \Phi^t \mathbf{Y}) \rho_{\mathbf{k}}(\mathbf{Y}) \, d\mathbf{Y} \qquad (173)$$

where $\mathbf{a}(\mathbf{X})$ is a function from the phase space to the periodic $f$-dimensional physical space which describes the jumps of the trajectories at the periodic boundaries. This Perron-Frobenius operator also admits Pollicott-Ruelle resonances and corresponding eigenstates

$$\hat{P}^t_{\mathbf{k}} |\phi_{\mathbf{k}n}\rangle = e^{s_n(\mathbf{k})t} |\phi_{\mathbf{k}n}\rangle \qquad \text{for} \qquad t > 0 \qquad (174)$$

depending on $\mathbf{k}$. In the case of diffusion, the leading resonance $s_0(\mathbf{k})$ gives the dispersion relation: $s_0(\mathbf{k}) = -Dk^2 + \mathcal{O}(k^4)$ [53,56]. This result shows the connection with the chaotic scattering approach were the leading Pollicott-Ruelle resonance or escape rate is $s_0 = -\gamma = -D(\pi/L)^2 + \mathcal{O}(L^{-3})$ so that we can identify the wavenumber of the corresponding hydrodynamical mode as $k = (\pi/L) + \mathcal{O}(L^{-2})$ [2].

For the multibaker model, the connection between the Pollicott-Ruelle resonances of the closed and scattering systems have been described in detail in Ref. [48] and the spectral decomposition Eq. (174) has been explicitly constructed in Ref. [53]. Tasaki and Gaspard [57] have further shown that a nonequilibrium steady state corresponding to a constant concentration gradient is obtained in the zero-wavenumber limit as $\partial_{\mathbf{k}}|\phi_{\mathbf{k}0}\rangle|_{\mathbf{k}=0}$ of the leading eigenstate of Eq. (174). Accordingly, such nonequilibrium steady states are also given by distributions. Moreover, Tasaki and Gaspard have proved for the multibaker model that the nonequilibrium steady states are described by singular measures having for support the full phase space and, nevertheless, exhibiting self-similar properties. A remarkable fact is that these self-similar properties have their origin in the fractal properties of the repeller and, in particular, of the escape-time functions of Eqs. (76) and (77), as shown by taking the large-system limit ($\mathcal{B} \to \mathcal{M}$) in order to connect the open and scattering configuration to the closed and infinitely extended configuration.

### B. Periodic-orbit theory

By using periodic-orbit theory [56,58,59], one can obtain in hyperbolic systems the characteristic equation for the resonances of the Perron-Frobenius operator in terms of the periodic orbits which thus appear to be directly related to the kinetics of the system. The periodic-orbit theory is based on the trace of the Perron-Frobenius operator which is given by

$$\text{tr } \hat{P}_{\mathbf{k}}^t = \sum_p T_p \sum_{r=1}^{\infty} e^{ir\mathbf{k}\cdot\mathbf{a}_p} \frac{\delta(t - rT_p)}{|\det(\mathbf{1} - \mathbf{M}_p^r)|} \quad (175)$$

where the first sum extends over the prime periodic orbits $\{p\}$ and the second sum over their repetitions. $T_p$ is the prime period of $p$, $\mathbf{M}_p$ the linearized Poincaré mapping in a surface of section transverse to the periodic orbit, $\mathbf{a}_p$ is a spatial vector representing the distance travelled in the full phase space when the orbit closes on itself in the reduced phase space. The Laplace transform of this trace can be expressed as

$$\int_0^{\infty} e^{-st} \text{ tr } \hat{P}_{\mathbf{k}}^t \, dt = \frac{\partial}{\partial s} \ln Z(s) \quad (176)$$

in terms of a Selberg-Smale zeta function defined by a product over the prime periodic orbits

$$Z(s) = \prod_p \prod_{m_1 \ldots m_L = 0}^{\infty} \left[ 1 - \frac{\exp(i\mathbf{k}\cdot\mathbf{a}_p - sT_p)}{\prod_{i=1}^L |\Lambda_p^{(i)}|\Lambda_p^{(i)\ m_i}} \right]^{(m_1+1)\ldots(m_L+1)} \quad (177)$$

Here the quantities $\Lambda_p^{(i)}$ are the expanding eigenvalues of the matrix $\mathbf{M}_p$, and $L$ is the total number of expanding directions. The zeros of the Selberg-Smale zeta function are the Pollicott-Ruelle resonances: $Z[s_n(\mathbf{k})] = 0$. Zeros are obtained after transforming the infinite product into a series by cycle expansion. In this way, a formula has been obtained [56] which expresses the diffusion coefficient in terms of the Lyapunov exponents of each periodic orbit, their periods, and the vectors $\mathbf{a}_p$

$$D = \frac{1}{4} \frac{\sum_{n=0}^{\infty} \sum_{p_1 \ldots p_n} (-)^n (\mathbf{a}_{p_1} + \cdots + \mathbf{a}_{p_n})^2 / |\Lambda_{p_1} \cdots \Lambda_{p_n}|}{\sum_{n=0}^{\infty} \sum_{p_1 \ldots p_n} (-)^n (T_{p_1} + \cdots + T_{p_n})^2 / |\Lambda_{p_1} \cdots \Lambda_{p_n}|} \quad (178)$$

for a two-degree-of-freedom system for which there is a single unstable direction ($L = 1$).

The periodic-orbit method has been applied not only to the study of diffusion in the Lorentz gas [56] and in other models [60], but also to the calculation of the Pollicott-Ruelle resonances of the three-disk scatterers [54] as well as of open and closed configurations of the multibaker [48]. The method is particularly interesting to obtain resonances beyond the escape rate which are difficult to obtain by direct simulation methods. However, the method is very demanding because a large number of periodic orbits may be required as it is the case in the Lorentz gas.

We conclude that there exist direct connections between the theory described in the present paper and the periodic orbit theory of the Perron-Frobenius operator. In both theories, the very same escape rate can be obtained as well as the other characteristic quantities of chaos such as the Lyapunov exponents and the partial Hausdorff dimension, as described elsewhere [16,17].

Recently Morriss, Rondoni, and Cohen have applied periodic-orbit theory to obtain the equilibrium pressure exerted by moving particles on the scatterers in a periodic, triangular Lorentz gas [61]. By carrying out computer simulations these authors were able to show that within the numerical errors, the equilibrium-ensemble measures are equivalent to dynamical measures based upon the stretching factors $\Lambda_p^{(i)}$ whose logarithms are the positive Lyapunov exponents for the corresponding periodic orbit. This result provides a further support to the ideas of Sinai, Ruelle, and Bowen on the dynamical foundations of the equilibrium measures.

### C. Thermostatted-system approach

In this approach, the system is subjected to some external field such as an electric field and to a special thermostatting force which maintains a constant kinetic energy [62–67]. These systems are deterministic and the dynamics is given by a set of ordinary differential equations, $\dot{\mathbf{X}} = \mathbf{F}(\mathbf{X})$, which is time-reversal symmetric under $\Theta$: $(\mathbf{q}, \mathbf{p}, t) \to (\mathbf{q}, -\mathbf{p}, -t)$. However, contrary to Hamiltonian systems where phase-space volumes are preserved $\mathrm{div}\mathbf{F} = 0$, the dynamics of thermostatted systems do not preserve phase-space volume but

$$\begin{aligned} \mathrm{div}\,\mathbf{F} &< 0 \quad \text{for} \quad \mathbf{X} \in \mathcal{M}_+ \\ \mathrm{div}\,\mathbf{F} &> 0 \quad \text{for} \quad \mathbf{X} \in \mathcal{M}_- \end{aligned} \quad (179)$$

so that the phase space is composed of two parts $\mathcal{M} = \mathcal{M}_+ \cup \mathcal{M}_-$ which are mapped onto each other by time reversal $\Theta \mathcal{M}_+ = \mathcal{M}_-$. The divergence of the vector field appears to be directly related to the transport properties by the way the thermostatting force is constructed. Moreover, because of the assumed coupling to a thermostat, an entropy production is defined in this approach.

As a consequence of Eq. (179), the domain $\mathcal{M}_+$ attracts the future of the trajectories and contains an attractor $\mathcal{A}_+$. By time reversibility, the other domain $\mathcal{M}_-$ contains a repeller $\mathcal{A}_-$ which is repelling in every directions (contrary to the repellers $\mathcal{R}$ considered in the present paper which are of saddle type). Under certain conditions, the attractor $\mathcal{A}_+$ may be a chaotic and fractal attractor to which the thermodynamic formalism described here is of application as shown by Chernov et al. [66]. These authors show, among othert things, that the attractor is the support of a Sinai-Bowen-Ruelle invariant measure $\mu_{\mathcal{A}_+}$ for which the Pesin formula holds, and the escape rate vanishes, $\gamma = 0$.

Recently, Gallavotti and Cohen [68] have described in some detail the construction of the SRB measures for the attractor, using methods based on Markov partitions. In this approach, they were able to show that the SRB measures can be used to provide the theoretical foundations for, and to quantitively describe the numerically observed fluctuations of the thermostatted-system entropy production in a shearing fluid far from equilibrium. Here, too, one has a clear example of the utility of constructing dynamical measures to describe nonequilibrium phenomena.

For thermostatted systems, the transport coefficients are related to the sum of all the Lyapunov exponents, $\sum_i \lambda_i$, which is negative in the domain $\mathcal{M}_+$ and, therefore, on the attractor $\mathcal{A}_+$ because of Eq. (179), in contrast to our escape-rate formula Eq. (2) in which the transport coefficient is related to the difference between the sum of positive Lyapunov exponents and the KS entropy for trajectories on the repeller. Nevertheless, there appear to be important structural similarities between the two expressions for transport coefficients, which have been commented upon in the literature [69,70] since the two kinds of expressions are given by the ratios of dynamical quantities to the squares of external parameters - the system size in the escape-rate formula, and the external field strength in the thermostat formalism. In both approaches, the hydrodynamical scale is related to the kinetic and chaotic time scale by a mechanism of dimensional loss in phase space. However, the mechanism of dimensional loss turns out to be different. In the chaotic-scattering approach, the dimensional loss is due to the finiteness of the scatterer and to the escape resulting from the nonequilibrium boundary condition. On the other hand, the non-volume-preserving thermostatting force is at the origin of dimensional loss in the thermostatted-system approach. We might mention here also that a somewhat different approach to dimensional reduction in systems with shearing flows has been developed by Chernov and Lebowitz [71] based upon the construction of special boundary conditions which maintain the shearing flow. The connections between all of these methods still remain to be established.

In section IV, we have seen that fractal dimensions can also be defined for probabilistic systems. Similarly, the concept of KS entropy has been generalized into a concept of $\varepsilon$-entropy to characterize stochastic processes [46]. This observation shows that the concepts of fractal and of chaos are not restricted to deterministic systems. Accordingly, we can imagine that it may be possible to characterize a mechanism of dimensional loss even in stochastic systems which describe systems coupled to thermal baths and in which the bath degrees of freedom are taken into account by fluctuating forces along with dissipative forces as in Langevin processes. In this larger perspective, we think that connections may exist between the descriptions of the transport and dynamical properties of thermostatted systems, and those with stochastic boundary conditions, with the chaotic-scattering approach described in this paper.

## VI. CONCLUSION

In this paper, we have reviewed the derivation of the escape-rate formula for a variety of systems of interest for studies of transport processes in fluids. The derivation is based on the local stretching rates which we introduced in Eq. (30) for smooth hyperbolic systems. Beside smooth hyperbolic systems, we have also considered the billiard systems and the cellular-automata lattice gases. The relation between the escape-rate formalism and transport coefficients was discussed in I, and a number of examples of the use of this formalism have appeared in the literature recently [16–19,48]. It is of course essential to explore a number of specific cases in order to develop some intuition about the properties of the fractal repellers that control the escape-rate, and to study cases where transport is anomalous in some way in order to determine the limitations of this approach.

There are other approaches to expressing transport coefficients in terms of the dynamical properties of the system under consideration. Such approaches have involved cycle expansions of dynamical zeta functions, or other properties of the zeta functions, the effects of Gaussian thermostats on the Lyapunov exponents of the system, as well as studies of the nature of the hydrodynamic modes that describe transport in the system. The connections between these various methods are only partially understood today. Clarifying these connections as well as establishing connections with more traditional approaches to transport theory based upon kinetic equations or Green-Kubo formulae remains a fruitful area for further study.

For a deeper understanding of the escape-rate formalism, it is essential to find methods of determining the KS entropy for trajectories on the repeller in a way that does not rely on the escape-rate formula. For certain uniformly hyperbolic systems, this can be accomplished by taking advantage of the equality of the KS and topological entropies, since the topological entropy can be determined in terms of the eigenvalues of the topological transition matrix [48,72]. However, to have more general methods, we will need to develop a deeper understanding of the dynamics on the fractal repeller with a possible development of a theory of Markov partitions for such trajectories. Another promising approach is based on the "Maryland" algorithm [73]. Here, for two-degree-of-freedom systems, one calculates the Hausdorff codimension of the repeller by studying the intervals of continuity of the escape-time function such as that illustrated in Fig. 4. One can then show that the Hausdorff codimension is related to the KS entropy in the large-system limit. This method has been applied to the determination of the diffusion coefficient for the Lorentz gas on a triangular lattice [16,17]. However, the generalization of such considerations to higher-dimensional fractals is still in its infancy [74,75].

Finally, we mention again that the Sinai method for determining the dynamical properties of billiard systems can be used as the basis of a more rigorous kinetic theory of gases, based not on the assumption of molecular chaos, but based rather on the dynamical chaos of billiard systems. In this direction, Bunimovich and Spohn have been able to provide a rigorous proof of the existence and positivity for the shear and bulk viscosities of a periodic billiard fluid with two hard disks per unit cell [76]. Moreover, Simányi and Szász [77] have shown that it may indeed be possible to complete Sinai's arguments for a rigorous proof of the ergodic behavior for a system of hard spheres.

## ACKNOWLEDGEMENTS

The authors would like to thank Matthieu Ernst, Henk van Beijeren, Arnulf Latz, Nelson Markley, E.G.D. Cohen, and Donald Jacobs for many enlightening discussions on a number of topics discussed here. J.R.D. wishes to thank in particular Prof. Ernst for his help on a number of issues related to cellular-automata lattice gases discussed in section IV, including a simple proof of the equivalence of the two methods described at the end of that section, and Prof. Cohen for informing him of recent work on SRB measures for systems in nonequilibrium stationary states and on dynamical calculations of equilibrium properties. P.G. would like to thank Prof. G. Nicolis for support and encouragement as well as the National Fund for Scientific Research (F. N. R. S. Belgium) and the "Communauté française de Belgique" (ARC contract No 93/98-166) for financial support. J.R.D. also acknowledges support from the National Science Foundation under grant PHY-93-21312.

**Figure captions**

Fig. 1. Schematic representation of the stable $W_s(\mathbf{X})$ and unstable $W_u(\mathbf{X})$ manifolds of the trajectory from initial condition $\mathbf{X}$. $\Sigma$ is a surface of section transverse to the orbit. The invariant manifolds are shown on the respective sides of $\Sigma$ which are used for their construction.

Fig. 2. Schematic behavior of the multivariate pressure function $P(\beta_1, \beta_2)$ for a three-degree-of-freedom system.

Fig. 3. Schematic behavior of the pressure function in the cases of: (a) a closed hyperbolic system; (b) an open hyperbolic system; (c) a closed nonhyperbolic system; (d) an open nonhyperbolic system. $\gamma$ denotes the escape rate and $\gamma_{\text{eff}}$ are the effective escape rate; $u$ the sum of the mean positive Lyapunov exponents; $h_{\text{KS}}$ the KS entropy per unit time; $h_{\text{top}}$ the topological entropy per unit time.

Fig. 4. Typical behavior of the escape-time function versus the initial condition $x_0$, here for the 1D logistic map $x_{n+1} = \mu x_n (1 - x_n)$ with $\mu = 4.01$.

Fig. 5. Examples of marginally unstable periodic orbits with zero Lyapunov exponents in the cases of (a) the stadium billiard, (b) the Sinai billiard, (c) the hard-sphere gas with periodic boundary conditions, (d) the hard-sphere gas in a rectangular box.

Fig. 6. Geometry of an elastic collision. $\mathbf{n}$ is the vector normal to the hypersurface of the billiard. $\mathbf{u}^{(\pm)}$ are respectively the velocities before and after the collision. $\Im$ is the tangent subspace. $\Im^{(\pm)}$ are the subspace perpendicular to the ingoing and outgoing velocities. $\delta\mathbf{q}^{(\pm)}$ are the infinitesimal perturbations in position before and after collision. $\delta\mathbf{q}$ is the infinitesimal perturbation of the impact point $\mathbf{q}$. The curvature of the collision hypersurface is not represented for the following reason. The hypersurface curvature plays a role only in the infinitesimal perturbations in velocities $\delta\mathbf{u}^{(\pm)}$ which are straightforwardly derived from the collision rule (112) by analysis as done in Eqs. (134)-(136). This analytical derivation does not particularly require a geometrical visualization.

Fig. 7. Expanding and contracting horospheres of a billiard. These horospheres are geometrical representations of the local unstable and stable manifolds, $W_u^{(\text{loc})}(\mathbf{X})$ and $W_s^{(\text{loc})}(\mathbf{X})$, at the phase-space point $\mathbf{X} = (\mathbf{q}, \mathbf{u})$. (a) Backward trajectory determining the expanding horosphere. (b) Forward trajectory determining the contracting horosphere.

Fig. 8. Hard-sphere gas in a box with convex, defocusing walls: (a) Each wall is composed of a portion of a single large sphere; (b) of the portions of many spheres modeling the atoms of the wall.